\documentclass[prd,superscriptaddress,showpacs,onecolumn]{revtex4-2}
\usepackage{amssymb}
\usepackage{bm}
\usepackage{amsfonts}
\usepackage{latexsym}
\usepackage[latin1]{inputenc}
\usepackage{graphicx}
\usepackage{amsmath}
\usepackage{palatino}
\usepackage{mathpazo}
\usepackage{textcomp}
\usepackage{float}
\usepackage{booktabs}
\usepackage{dcolumn}
\usepackage{hyperref}
\usepackage{amsmath}
\usepackage{xcolor}
\usepackage{lipsum}
\usepackage{orcidlink}
\usepackage[caption=false]{subfig}
\usepackage{commath}

\hypersetup{colorlinks,citecolor=blue}
\def\jnl@style{\it}
\def\aaref@jnl#1{{\jnl@style#1}}
\def\aaref@jnl#1{{\jnl@style#1}}
\def\aj{\aaref@jnl{AJ}}
\def\apj{\aaref@jnl{ApJ}}
\def\apjl{\aaref@jnl{ApJ}}
\def\apjs{\aaref@jnl{ApJS}}
\def\apss{\aaref@jnl{Ap\&SS}}
\def\aap{\aaref@jnl{A\&A}}
\def\aapr{\aaref@jnl{A\&A~Rev.}}
\def\aaps{\aaref@jnl{A\&AS}}
\def\mnras{\aaref@jnl{Mon.~Not.~Roy.~Astron.~Soc.}}
\def\prd{\aaref@jnl{Phys.~Rev.~D}}
\def\prc{\aaref@jnl{Phys.~Rev.~C}}
\def\prl{\aaref@jnl{Phys.~Rev.~Lett.}}
\def\qjras{\aaref@jnl{QJRAS}}
\def\skytel{\aaref@jnl{S\&T}}
\def\ssr{\aaref@jnl{Space~Sci.~Rev.}}
\def\zap{\aaref@jnl{ZAp}}
\def\nat{\aaref@jnl{Nature}}
\def\aplett{\aaref@jnl{Astrophys.~Lett.}}
\def\apspr{\aaref@jnl{Astrophys.~Space~Phys.~Res.}}
\def\physrep{\aaref@jnl{Phys.~Rep.}}
\def\physscr{\aaref@jnl{Phys.~Scr}}
\def\commat{\aaref@jnl{Comm.~Math.~Phys.}}
\def\science{\aaref@jnl{Science}}
\def\cqg{\aaref@jnl{Classical Quant.~Grav.}}
\def\jpcs{\aaref@jnl{JPCS}}
\def\ijmpd{\aaref@jnl{Int.~J.~Mod.~Phys.~D}}
\def\grg{\aaref@jnl{Gen.~Relat.~Gravit.}}
\def\rpp{\aaref@jnl{Rep.~Prog.~Phys.}}
\def\npa{\aaref@jnl{Nucl.~Phys.~A}}
\def\lrr{\aaref@jnl{Living Rev.~Rel.}}
\def\jcap{\aaref@jnl{J.~Cosmology Astropart.~Phys.}}
\def\rmp{\aaref@jnl{Rev.~Mod.~Phys.}}
\def\epjc{\aaref@jnl{Eur.~Phys.~J.~C}}

\allowdisplaybreaks[1]

\begin{document}

\noindent Phys. Dark Universe {\bf 32}, 100820 (2021) \hfill  \url{https://doi.org/10.1016/j.dark.2021.100820}

\title{Cosmic acceleration with bulk viscosity in modified $f(Q)$ gravity}
\author{Raja Solanki\orcidlink{-}}
\email{rajasolanki8268@gmail.com}
\affiliation{Department of Mathematics, Birla Institute of Technology and
Science-Pilani,\\ Hyderabad Campus, Hyderabad-500078, India.}
\author{S. K. J. Pacif\orcidlink{0000-0003-0951-414X}}
\email{shibesh.math@gmail.com}
\affiliation{Department of Mathematics, School of Advanced Sciences, Vellore
Institute of Technology, Vellore 632014, Tamil Nadu, India.}
\author{Abhishek Parida}
\email{abhishekparida22@gmail.com}
\affiliation{Centre for Theoretical Physics, Jamia Milia Islamia, New
Delhi-110025, India.}
\author{P.K. Sahoo\orcidlink{0000-0003-2130-8832}}
\email{pksahoo@hyderabad.bits-pilani.ac.in}
\affiliation{Department of Mathematics, Birla Institute of Technology and
Science-Pilani,\\ Hyderabad Campus, Hyderabad-500078, India.}
\date{\today }

\begin{abstract}
In this article, we have investigated the role of bulk viscosity to study
the accelerated expansion of the universe in the framework of modified $f(Q)$
gravity. The gravitational action in this modified gravity theory has the
form $f(Q)$, where $Q$ denote the non-metricity scalar. In the present
manuscript, we have considered a bulk viscous matter-dominated cosmological
model with the bulk viscosity coefficient of the form $\xi =\xi _{0}+\xi
_{1}H+\xi _{2}\left( \frac{\dot{H}}{H}+H\right) $ which is proportional to
the velocity and acceleration of the expanding universe. Two sets of
limiting conditions on the bulk viscous parameters $\xi _{0},$ $\xi _{1},$ $%
\xi _{2}$ and model parameter $\alpha $ arose here out of which one
condition favours the present scenario of cosmic acceleration with a phase
transition and corresponds to the universe with a Big Bang origin. Moreover,
we have discussed the cosmological behaviour of some geometrical parameters.
Then, we have obtained the best fitting values of the model parameters $\xi
_{0},$ $\xi _{1},$ $\xi _{2}$ and $\alpha $ by constraining our model with
updated Hubble datasets consisting of $57$ data points and recently released
Pantheon datasets consisting of $1048$ data points which show that our
obtained model has good compatibility with observations. Further, we
have also included the Baryon Acoustic Oscillation (BAO) datasets of six
data points with the Hubble \& Pantheon datasets and obtained
slightly different values of the model parameters. Finally, we have
analyzed our model with the statefinder diagnostic analysis and found some
interesting results and are discussed in details.
\end{abstract}

\maketitle

\color{black}

\section{Introduction}\label{sec1}

In the year 1915, Albert Einstein proposed the General Theory of Relativity
(GR). In the Solar System tests, GR is extremely successful as yet, but GR
does not give the final word to all gravity occurrences so far. Over the
last two decades, a convergence of several cosmological observations
indicate that our Universe is going through a period of accelerated
expansion. The observational data on type \textbf{I}a Supernovae have shown
that the current universe is accelerating \cite{Riess,Perlmutter}. The
observational evidence such as Baryon Acoustic Oscillations (BAO) \cite%
{D.J.,W.J.}, large scale structure \cite{T.Koivisto,S.F.}, galaxy redshift
survey \cite{C.Fedeli} and Cosmic Microwave Background Radiations (CMBR) 
\cite{R.R.,Z.Y.} strongly support this accelerating expansion. The observed
late-time acceleration of the Universe is one of the premier mysteries of
theoretical physics and the responsible mechanism for this accelerating
expansion is still an open question. Plenty of models have been proposed in
the literature to describe this recent acceleration. Basically, there are
two approaches to interpret this recent acceleration of the universe. The
first approach is the assumption of the existence of mysterious force with
high negative pressure so-called dark energy (DE) as responsible for the
current acceleration of the universe. The simplest candidate for dark energy
is the cosmological constant $\Lambda$ (or vacuum energy) i.e. the fluid
responsible for such an effective negative pressure with constant energy
density. This model is characterized by the constant equation of state (EoS)
parameter $\omega_\Lambda = -1 $ and known as $\Lambda$CDM model \cite%
{S.Weinberg,Carroll}. Even though this model agrees considerably well with
observational data, it is faced with some strong problems. Of these, the two
major drawbacks are the cosmic coincidence problem and the fine-tuning
problem \cite{E.J.}. In today's universe, the density of dark energy and the
density of non-relativistic matter happen to be the same order of magnitude,
even though their evolution is different. This observed coincidence between
the densities referred to as cosmic coincidence problem, while the fine
tuning problem refers to the inconsistency between the observed value and
theoretically predicted value of the cosmological constant. To overcome
these problems time-varying dark energy models have been proposed in the
literature like quintessence \cite{Carroll-2,Y.Fujii}, k-essence \cite%
{T.Chiba,C.Arm.} and perfect fluid models (like the Chaplygin gas model) 
\cite{M.C.,A.Y.}. The second approach to explain the current acceleration of
the universe is to modify spacetime's geometry. We can do this by modifying
the left-hand side of the Einstein equation. Modified theories of gravity
are the geometrical generalizations of Einstein's general theory of
relativity in which the cosmic acceleration can be achieved by modifying the
Einstein-Hilbert action of GR. Recently, modified theories of gravity have
attracted the interest of cosmologists for understanding the role of dark
energy. In modified gravity, the origin of dark energy is recognized as a
modification of gravity. A lot of research reveals that the modified
theories of gravity can explain both early and late time acceleration of the
universe. Hence, there are plenty of motivations to discover theories beyond
the standard formulation of GR. There are several modified theories have
been proposed in the literature like $\:f(R)\:$ theory \cite%
{Hans,A.A.,S.Nojiri}, $\:f(T)\:$ theory \cite{R.F.,E.V.,K.B.}, $f(T,B)$
theory \cite{Sebastian}, $\:f(R,T)\:$ theory \cite{Harko,Hamid}, $f(Q,T)$
theory \cite{Yixin,Simran}, $\:f(G)\:$ theory \cite{S.Nojiri-2}, $f(R,G)$
theory \cite{E.E.,K.B.-2}, etc. Nowadays, $f(Q)$ theories of gravity have
been extremely investigated. The symmetric teleparallel gravity or $f(Q)$
gravity was introduce by J.B. Jim\'enez et al. \cite{J.B.}. The $f(Q)$
theory is also an alternative theory for GR like teleparallel gravity. In
symmetric teleparallel gravity gravitational interactions are described by
the non-metricity $Q$. Recently, there are several studies done in $f(Q)$
gravity. T. Harko studied the extension of symmetric teleparallel gravity 
\cite{Harko-2}. S. Mandal studied energy conditions in $f(Q)$ gravity and 
also did a comparative study between $f(Q)$ gravity and $\Lambda$CDM 
\cite{Sanjay}. Moreover, they used the cosmographic idea to constrain the
Lagrangian function $f(Q)$ using the latest pantheon data \cite{Sanjay2}. An
interesting investigation on $f(Q)$ gravity was done by Noemi, where he
explored the signatures of non-metricity gravity in its' fundamental level 
\cite{Noemi/2021}.

Earlier, to study inflationary epoch in the early universe bulk viscosity
has been proposed in the literature without any requirement of dark energy 
\cite{T.P.,I.W.}. Hence, it is very natural to expect that the bulk
viscosity can be responsible for the current accelerated expansion of the
universe. Nowadays, several authors are attempted to explain the late-time
acceleration via bulk viscosity without any dark energy constituent or
cosmological constant \cite{Athira,Mohan,Simran-2,G.C.,J.C.,A.Av.}.
Theoretically, deviations that occur from the local thermodynamic stability
can originate the bulk viscosity but a detailed mechanism for the formation
of bulk viscosity is still not achievable \cite{W.Z.}. In cosmology, when
the matter content of the universe expands or contract too fast as a
cosmological fluid then the effective pressure is generated to bring back
the system to its thermal stability. The bulk viscosity is the manifestation
of such an effective pressure \cite{J.R.,H.O.}.

In cosmology, there is two main formalism for the description of bulk
viscosity. The first one is the non-casual theory, where the deviation of
only first-order is considered and one can find that the heat flow and
viscosity propagate with infinite speed while in the second one i.e. the
casual theory it propagates with finite speed. In the year 1940, Eckart
proposed the non-casual theory \cite{C.E.}. Later, Lifshitz and Landau gave
a similar theory \cite{L.D.}. The casual theory was developed by Israel,
Hiscock and Stewart. In this theory, second-order deviation from equilibrium
is considered \cite{W.I.,W.I.-2,W.I.-3,W.A.,W.A.-2}. Moreover, Eckart theory
can be acquired from it as a first-order approximation. Hence, Eckart's
theory is a good approximation to the Israel theory in the limit of
vanishing relaxation time. To analyze the late acceleration of the universe,
the casual theory of bulk viscosity has been used. Cataldo et al. have
investigated the late time acceleration using the casual theory \cite%
{M.Cataldo}. Basically, they used an ansatz for the Hubble parameter
(inspired by the Eckart theory) and they have shown the transition of the
universe from the big rip singularity to the phantom behavior.

The expansion process of an accelerating universe is a collection of states
that lose their thermal stability in a small fragment of time \cite{A.Av.-2}%
. Hence, it is quite natural to consider the existence of bulk viscosity
coefficient to describe the expansion of the universe. The accelerated
expansion scenario of the universe (the mean stage of low redshift) can be
justified by the geometrical modification in Einstein's equation. Also,
without any requirement of cosmological constant bulk viscosity can generate
an acceleration. It contributes to the pressure term and applies additional
pressure to drive the acceleration \cite{S.Od.}. C. P. Singh and Pankaj
Kumar has investigated the role of bulk viscosity in modified $f(R,T)$
theories of gravity \cite{Singh}. S. Davood has investigated the effect of
bulk viscous matter in modified $f(T)$ theories of gravity \cite{S.Davood}.

In this work, we have focused on studying the cosmic acceleration of
the universe in $f(Q)$ gravity with the presence of bulk viscous fluid. The
motivation of working in the non-metricity $f(Q)$ gravity is that in this
framework, the motion equations are in the second-order, which is easy to
solve. In f(R) gravity, an extra scalar mode appears because the model is
the higher derivative theory as the Ricci scalar includes the second-order
derivatives of the metric tensor. This scalar mode generates additional
force, and it is often inconsistent with the Newton law observations and
also for a density of a canonical scalar field $\phi$; the non-minimal
coupling between geometry and the matter Lagrangian produces an additional
kinetic term which is not an agreement with the stable Horndeski class \cite%
{Olmo/2015}. Nevertheless, the non-metricity formalism overcomes the above
problems, which are induced by the higher-order theory. 

In this article, we analyze the matter-dominated FLRW model in the framework
of modified $f(Q)$ theories of gravity and study the role of bulk viscosity
in explaining the late-time acceleration of the universe. The outline of the
present article is as follows. In Sec. \ref{sec2} we present the field
equation formalism in $f(Q)$ gravity. In Sec. \ref{sec3} we describe the
FLRW universe dominated with bulk viscous matter and also we derive the
expression for the Hubble parameter. In Sec. \ref{sec4} we derive the scale
factor and found two sets of limiting conditions on the coefficients of bulk
viscosity which corresponds to the universe which begins with a Big Bang and
then making a transition from deceleration phase to the acceleration phase.
In Sec. \ref{sec5} we show the evolution of deceleration parameter $q$. In
Sec. \ref{sec6} we have constrained the model parameters by using Hubble
data and Pantheon data sets. In Sec. \ref{sec7} we adopt the statefinder
diagnostic pair to differentiate present bulk viscous model with other
models of dark energy. Finally, in the last section Sec. \ref{sec8} we
briefly discuss our conclusions.

\section{Motion Equations in $f(Q)$ gravity}\label{sec2}

The action in a universe governed by $f(Q)$ gravity reads 
\begin{equation}
S=\int {\frac{1}{2}f(Q)\sqrt{-g}d^{4}x}+\int {L_{m}\sqrt{-g}d^{4}x}\text{{,}}
\label{2a}
\end{equation}%
where $f(Q)$ is an arbitrary function of the nonmetricity $Q$, $g$ is the
determinant of the metric $g_{\mu \nu }$ and $L_{m}$ is the matter
Lagrangian density.

The nonmetricity tensor is defined as 
\begin{equation}
Q_{\lambda \mu \nu }=\nabla _{\lambda }g_{\mu \nu }  \label{2b}
\end{equation}%
and its two traces are given below 
\begin{equation}
Q_{\alpha }={Q_{\alpha }^{\mu}}_{\mu} \ \ \text{and} \ \ \tilde{Q}%
_{\alpha}=Q^{\mu }_{\alpha \mu }\text{.}  \label{2c}
\end{equation}%
Moreover, the superpotential tensor is given by 
\begin{equation}
4P^{\lambda }_{\mu \nu }=-Q^{\lambda }_{\mu\nu}+2Q_{({\mu^{^{\lambda}}}{\nu}%
)}+(Q^{\lambda }-\tilde{Q}^{\lambda })g_{\mu \nu }-\delta^{\mu}_{({%
\lambda^{^{Q}}}\nu)}\text{.}  \label{2d}
\end{equation}%
Hence, the trace of nonmetricity tensor can be obtained as 
\begin{equation}
Q=-Q_{\lambda \mu \nu }P^{\lambda \mu \nu }\text{.}  \label{2e}
\end{equation}%
Now, the definition of the energy momentum tensor for the matter is 
\begin{equation}
T_{\mu \nu }=\frac{-2}{\sqrt{-g}}\frac{\delta (\sqrt{-g}L_{m})}{\delta
g^{\mu \nu }}\text{.}  \label{2f}
\end{equation}%
For notational simplicity, we define $f_{Q}=\frac{df}{dQ}$

Varying the action \eqref{2a} with respect to the metric, the gravitational
field equation obtained is given below 

\begin{equation}\label{2g}
\frac{2}{\sqrt{-g}}\nabla_\lambda (\sqrt{-g}f_Q P^\lambda_{\mu\nu}) + \frac{1}{2}g_{\mu\nu}f+f_Q(P_{\mu\lambda\beta}Q_\nu^{\lambda\beta} - 2Q_{\lambda\beta\mu}P^{\lambda\beta}_\nu) = -T_{\mu\nu}
\end{equation}

Furthermore, by varying the action \eqref{2a} with respect to the
connection, one can find the following result 
\begin{equation}
\nabla _{\mu }\nabla _{\nu }(\sqrt{-g}f_{Q}P^{\mu \nu}_{\lambda })=0\text{.}
\label{2h}
\end{equation}

\section{FLRW universe dominated with bulk viscous matter}\label{sec3}

We consider that the universe is described by the spatially flat
Friedmann-Lemaitre-Robertson-Walker(FLRW) line element 
\begin{equation}
ds^{2}=-dt^{2}+a^{2}(t)[dx^{2}+dy^{2}+dz^{2}]\text{.}  \label{3a}
\end{equation}%
Here, $a(t)$ is the scale factor of the universe dominated with bulk viscous
matter. The trace of nonmetricity tensor with respect to line element given
by \eqref{3a} is 
\begin{equation}
Q=6H^{2}\text{.}  \label{3b}
\end{equation}%
For a bulk viscous fluid, described by its effective pressure $\bar{p}$ and
the energy density $\rho $, the energy-momentum tensor takes the form 
\begin{equation}
T_{\mu \nu }=(\rho +\bar{p})u_{\mu }u_{\nu }+\bar{p}h_{\mu \nu }\text{,}
\label{3c}
\end{equation}%
where $h_{\mu \nu }=g_{\mu \nu }+u_{\mu }u_{\nu }$ and $\bar{p}=p-3\xi H$.
Here $\xi $ is the coefficient of bulk viscosity which can be a function of
Hubble parameter and its derivative and the components of four-velocity $%
u^{\mu }$ are $u^{\mu }=(1,0)$ and $p$ is the normal pressure which is 0 for
non-relativistic matter.

The Friedmann equations describing the universe dominated with bulk viscous
matter are 
\begin{equation}
3H^{2}=\frac{1}{2f_{Q}}\left( -\rho +\frac{f}{2}\right)  \label{3d}
\end{equation}%
and 
\begin{equation}
\dot{H}+3H^{2}+\frac{\dot{f_{Q}}}{f_{Q}}H=\frac{1}{2f_{Q}}\left( \bar{p}+%
\frac{f}{2}\right) \text{.}  \label{3e}
\end{equation}%
In an accelerated expanding universe, the coefficient of viscosity should
depend on velocity and acceleration. In this paper, we consider a time
dependent bulk viscosity of the form \cite{J.Ren} 
\begin{equation}
\xi =\xi _{0}+\xi _{1}\left( \frac{\dot{a}}{a}\right) +\xi _{2}\left( \frac{{%
\ddot{a}}}{\dot{a}}\right) =\xi _{0}+\xi _{1}H+\xi _{2}\left( \frac{\dot{H}}{%
H}+H\right) \text{.}  \label{3f}
\end{equation}%
It is a linear combination of three terms, first one is a constant, second
one is proportional to the Hubble parameter, which indicates the dependence
of the viscosity on speed, and the third one is proportional to the $\frac{%
\ddot{a}}{\dot{a}}$, indicating the dependence of the bulk viscosity on
acceleration.

In this paper, we consider the following functional form of $f(Q)$ 
\begin{equation}
f(Q)=\alpha Q,\ \ \ \alpha \neq 0\text{.}  \label{3g}
\end{equation}%
Then, for this particular choice of the function, the field equation becomes 
\begin{equation}
\rho =-3\alpha H^{2}  \label{3h}
\end{equation}%
and 
\begin{equation}
\bar{p}=2\alpha \dot{H}+3\alpha H^{2}\text{.}  \label{3i}
\end{equation}%
As we are concerned with late-time acceleration, we have considered the
non-relativistic matter dominates the universe. From the Friedmann equation %
\eqref{3i} and equation \eqref{3f}, we have first-order differential
equation for the Hubble parameter by replacing $\frac{d}{dt}$ with $\frac{d}{%
dln(a)}$ via $\frac{d}{dt}=H\frac{d}{dln(a)}$

\begin{equation}
\frac{dH}{dln(a)}+\left( \frac{3\alpha +3\xi _{1}+3\xi _{2}}{2\alpha +3\xi
_{2}}\right) H+\left( \frac{3\xi _{0}}{2\alpha +3\xi _{2}}\right) =0\text{.}
\label{3j}
\end{equation}

Now, we set 
\begin{equation}
3\xi _{0}=\bar{\xi _{0}}H_{0}, \ \ 3\xi _{1}=\bar{\xi _{1}}, \ \ 3\xi _{2}=%
\bar{\xi _{2}} \ \ \text{and} \ \ \bar{\xi}_{12}=\bar{\xi _{1}}+\bar{\xi _{2}%
}\text{,}  \label{3k}
\end{equation}%
where $H_{0}$ is present value of the Hubble parameter and $\bar{\xi}_{0},\
\ $ $\bar{\xi}_{1},\ \ $ $\bar{\xi}_{2}$ are the dimensionless bulk viscous
parameters, then by using above equation \eqref{3j} becomes

\begin{equation}
\frac{dH}{dln(a)}+\left( \frac{3\alpha +\bar{\xi}_{12}}{2\alpha +\bar{\xi
_{2}}}\right) H+\left( \frac{\bar{\xi _{0}}}{2\alpha +\bar{\xi _{2}}}\right)
H_{0}=0\text{.}  \label{3l}
\end{equation}

After integrating above equation we obtain the Hubble parameter as 
\begin{equation}
H(a)=H_{0}\left[ a^{-\left( \frac{3\alpha +\bar{\xi}_{12}}{2\alpha +\bar{\xi
_{2}}}\right) }\left( 1+\frac{\bar{\xi _{0}}}{3\alpha +\bar{\xi}_{12}}%
\right) -\frac{\bar{\xi _{0}}}{3\alpha +\bar{\xi}_{12}}\right] \text{.}
\label{3m}
\end{equation}

At $\bar{\xi}_{0}=\bar{\xi}_{1}=\bar{\xi}_{2}=0$, equation \eqref{3m}
becomes 
\begin{equation}
H=H_{0}a^{-\frac{3}{2}}\text{.}  \label{3n}
\end{equation}%
The equation \eqref{3n} gives the value of the Hubble parameter in case of
ordinary matter-dominated universe i.e. when all the bulk viscous parameters
are $0$. \newline
Now, by using the relation between redshift and scale factor i.e $a(t)=\frac{%
1}{1+z}$, in terms of redshift the Hubble parameter is given as 
\begin{equation}
H(z)=H_{0}\left[ (1+z)^{\left( \frac{3\alpha +\bar{\xi}_{12}}{2\alpha +\bar{%
\xi _{2}}}\right) }\left( 1+\frac{\bar{\xi _{0}}}{3\alpha +\bar{\xi}_{12}}%
\right) -\frac{\bar{\xi _{0}}}{3\alpha +\bar{\xi}_{12}}\right] \text{.}
\label{3o}
\end{equation}

\section{Scale factor}\label{sec4}

Now, using the definition of Hubble parameter, the equation \eqref{3m}
becomes 
\begin{equation}
\frac{1}{a}\frac{da}{dt}=H_{0}\left[ a^{-\left( \frac{3\alpha +\bar{\xi}_{12}%
}{2\alpha +\bar{\xi _{2}}}\right) }\left( 1+\frac{\bar{\xi _{0}}}{3\alpha +%
\bar{\xi}_{12}}\right) -\frac{\bar{\xi _{0}}}{3\alpha +\bar{\xi}_{12}}\right]
\text{.}  \label{4a}
\end{equation}%
On integrating the above equation we get the scale factor 
\begin{equation}
a(t)=\left[ \frac{3\alpha +\bar{\xi}_{12}+\bar{\xi}_{0}}{\bar{\xi}_{0}}%
-\left( \frac{3\alpha +\bar{\xi}_{12}}{\bar{\xi}_{0}}\right)
e^{-H_{0}(t-t_{0})\frac{\bar{\xi}_{0}}{2\alpha +\bar{\xi}_{2}}}\right] ^{%
\frac{2\alpha +\bar{\xi}_{2}}{3\alpha +\bar{\xi}_{12}}}\text{,}  \label{4b}
\end{equation}%
where $t_{0}$ is the present cosmic time.

Now, let $y=H_0(t-t_0) $ , then the second order derivative of $a(t) $ with
respect to $y$ is

\begin{equation}
\frac{d^2a}{dy^2} = \frac{e^{\frac{- \bar{\xi}_0 y}{2\alpha+ \bar{\xi}_2}}}{2\alpha+\bar{\xi}_2} \left(-(\bar{\xi}_0+ \bar{\xi}_{12}+3\alpha) + (2\alpha + \bar{\xi}_2) e^{\frac{- \bar{\xi}_0 y}{2\alpha+ \bar{\xi}_2}} \right)  
 \times \left[\frac{\bar{\xi}_0 + \bar{\xi}_{12} + 3\alpha - (3\alpha + \bar{\xi}_{12}) e^{\frac{- \bar{\xi}_0 y}{2\alpha + \bar{\xi}_2}}}{\bar{\xi}_0} \right]^{\frac{-2(2\alpha + \bar{\xi}_1)- \bar{\xi}_2}{3\alpha + \bar{\xi}_{12}}}\text{.}
\end{equation}

From the above expression it is clear that we have two limiting conditions
based on the values of $\bar{\xi}_{0},$ $\bar{\xi}_{1}$ and $\bar{\xi}_{2}$.
By equation \eqref{3h} and the fact that density of ordinary matter in the
universe is always positive, so we must have $\alpha <0$. Assuming, $\alpha
=-\bar{\alpha}$ where $\bar{\alpha}>0$, these two limiting conditions are

\begin{equation}
\bar{\xi}_{0}>0,\text{ }\bar{\xi}_{12}<3\bar{\alpha},\text{ }\bar{\xi}_{2}<2%
\bar{\alpha},\text{ }\bar{\xi}_{0}+\bar{\xi}_{12}<3\bar{\alpha}  \label{4c}
\end{equation}%
and 
\begin{equation}
\bar{\xi}_{0}<0,\text{ }\bar{\xi}_{12}>3\bar{\alpha},\text{ }\bar{\xi}_{2}>2%
\bar{\alpha},\text{ }\bar{\xi}_{0}+\bar{\xi}_{12}>3\bar{\alpha}\text{.}
\label{4d}
\end{equation}
These two limiting condition implies that the universe experienced a
deceleration phase at early times and then making a transition into the
accelerated phase in the latter times. If we place $\bar{\xi}_{0}+\bar{\xi}%
_{12}>3\bar{\alpha}$ and $\bar{\xi}_{0}+\bar{\xi}_{12}<3\bar{\alpha}$ in the
first and second limiting condition respectively, then the universe will
experience an everlasting accelerated expansion.

\section{Deceleration parameter}\label{sec5}

\begin{figure}[H]
\centering
\includegraphics[scale=0.71]{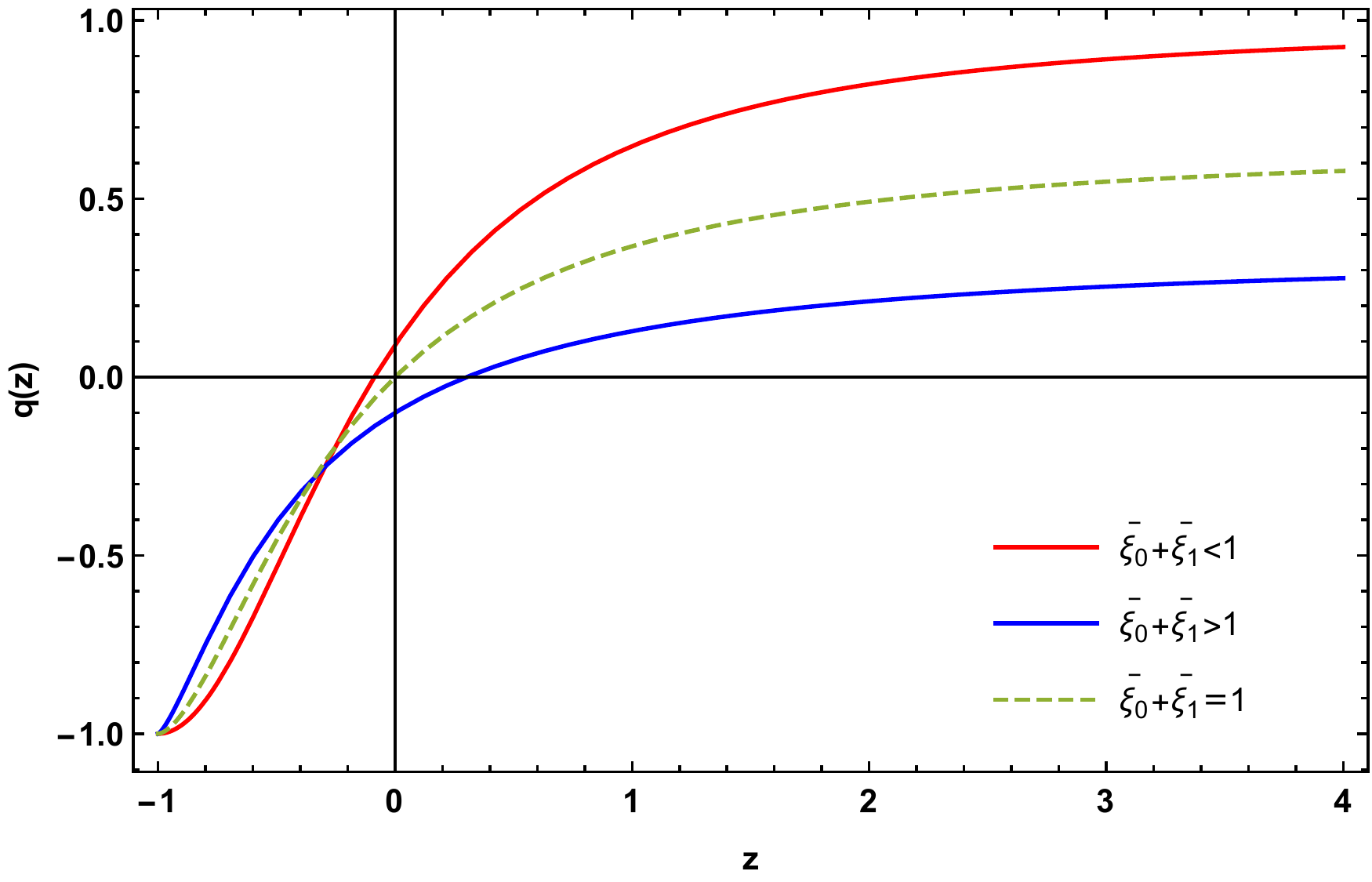}
\caption{Variation of the deceleration parameter with redshift $z$ for the
first limiting conditions $\bar{\protect\xi}_{0}>0, \bar{\protect\xi}%
_{12}<3, \bar{\protect\xi}_{2}<2, \bar{\protect\xi}_{0}+\bar{\protect\xi}%
_{12}<3$. Here, we took $\protect\alpha =-1$ i.e., $\bar{\protect\alpha}=1 $%
. $q$ enters the negative region in the recent past if $\bar{\protect\xi}%
_{0}+\bar{\protect\xi}_{1}>1$, at present if $\bar{\protect\xi}_{0}+\bar{%
\protect\xi}_{1}=1$ and in the future if $\bar{\protect\xi}_{0}+\bar{\protect%
\xi}_{1}<1$. For Red, Blue and Green plots the value of $(\bar{\protect\xi}%
_{0},\bar{\protect\xi}_{1},\bar{\protect\xi}_{2})$ are $%
(0.9,0.01,1),(0.45,0.65,1),(0.65,0.35,1)$ respectively. }
\label{f1}
\end{figure}

\begin{figure}[H]
\centering
\includegraphics[scale=0.71]{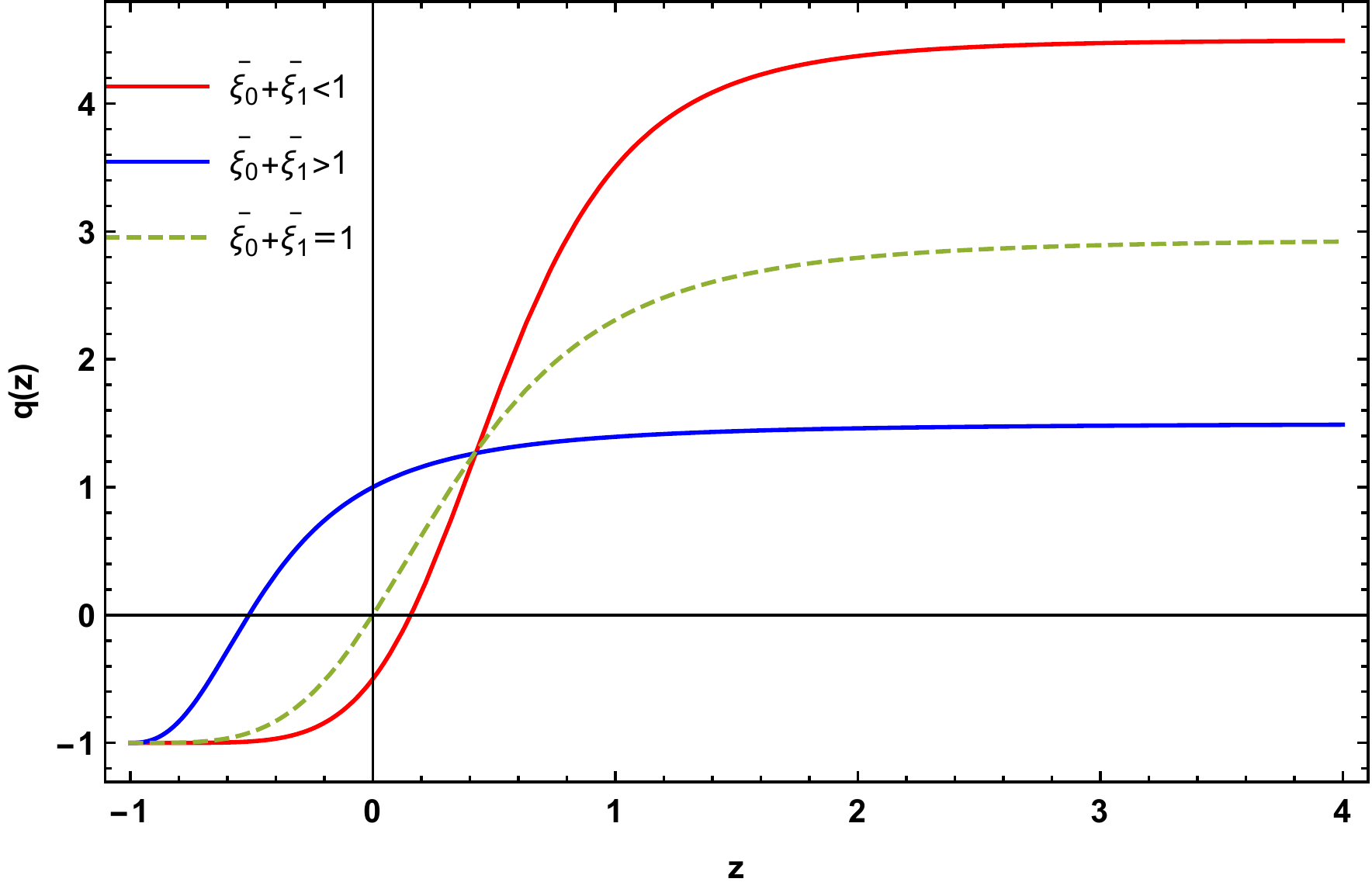}
\caption{Variation of the deceleration parameter with redshift $z$ for the
second limiting conditions $\bar{\protect\xi}_{0}<0, \bar{\protect\xi}%
_{12}>3, \bar{\protect\xi}_{2}>2, \bar{\protect\xi}_{0}+\bar{\protect\xi}%
_{12}>3$. Here, we took $\protect\alpha =-1$ i.e., $\bar{\protect\alpha}=1 $%
. $q$ enters the negative region in the recent past if $\bar{\protect\xi}%
_{0}+\bar{\protect\xi}_{1}<1$, at present if $\bar{\protect\xi}_{0}+\bar{%
\protect\xi}_{1}=1$ and in the future if $\bar{\protect\xi}_{0}+\bar{\protect%
\xi}_{1}>1$. For Red, Blue and Green plots the value of $(\bar{\protect\xi}%
_{0},\bar{\protect\xi}_{1},\bar{\protect\xi}_{2})$ are $%
(-0.5,1.45,2.1),(-0.5,2.5,3),(-0.5,1.5,2.17)$ respectively}
\label{f2}
\end{figure}

The deceleration parameter is defined as 
\begin{equation}
q=-\frac{a\ddot{a}}{\dot{a}^{2}}=-\frac{\ddot{a}}{a}\frac{1}{H^{2}}\text{.}
\label{5a}
\end{equation}%
From the Friedmann equation \eqref{3i} one can obtain 
\begin{equation}
\frac{\ddot{a}}{a}=-\frac{1}{2\alpha }\left[ \alpha H^{2}+3H\left( \xi
_{0}+\xi _{1}H+\xi _{2}\left( \frac{\dot{H}}{H}+H\right) \right) \right] 
\text{.}  \label{5b}
\end{equation}%
Using the equation \eqref{3k} the deceleration parameter becomes 
\begin{equation}
q=\frac{1}{2\alpha }\left[ \bar{\xi}_{0}\frac{H_{0}}{H}+\left( \bar{\xi}%
_{12}+\alpha \right) +\bar{\xi}_{2}\frac{\dot{H}}{H^{2}}\right] \text{.}
\label{5c}
\end{equation}
Now, using the value of Hubble parameter given by equation \eqref{3m} and
equations \eqref{3i}-\eqref{3l}, the equation \eqref{5c} becomes 
\begin{equation}
q(a)=\frac{1}{2\alpha +\bar{\xi}_{2}}\left[ \bar{\xi}_{1}+\alpha +\frac{\bar{%
\xi}_{0}}{a^{-\left( \frac{3\alpha +\bar{\xi}_{12}}{2\alpha +\bar{\xi}_{2}}%
\right) }\left[ 1+\frac{\bar{\xi}_{0}}{3\alpha +\bar{\xi}_{12}}\right] -%
\frac{\bar{\xi}_{0}}{3\alpha +\bar{\xi}_{12}}}\right] \text{.}  \label{5d}
\end{equation}
Now, by using relation between redshift and scale factor i.e., $a(t)=\frac{1%
}{1+z}$, in terms of redshift deceleration parameter is given as 

\begin{equation}\label{5e}
q(z)=\frac{1}{2\alpha + \bar{\xi}_2} \left[\bar{\xi}_1+\alpha + \frac{\bar{\xi}_0}{(1+z)^{\left(\frac{3\alpha+ \bar{\xi}_{12}}{2\alpha+ \bar{\xi}_2}\right)} \left[ 1+ \frac{\bar{\xi}_0}{3\alpha+ \bar{\xi}_{12}}\right] - \frac{\bar{\xi}_0}{3\alpha+\bar{\xi}_{12}}} \right]
\end{equation}
The present value of deceleration parameter i.e., the value of 
$q$ at $z=0$ or $a=1$ is, 
\begin{equation}
q_{0}=\frac{\alpha +\bar{\xi}_{0}+\bar{\xi}_{1}}{2\alpha +\bar{\xi}_{2}}%
\text{.}  \label{5f}
\end{equation}%
If the value of all bulk viscous parameters are 0, then the deceleration
parameter becomes $q=\frac{1}{2}$ which correspond to a matter dominated
decelerating universe with null bulk viscosity. For the two sets of limiting
conditions based on the dimensionless bulk viscous parameter, the variation
of deceleration parameter with respect to redshift $z$ can be plotted as
shown in Figs. \ref{f1} and \ref{f2}.


\begin{figure}[H]
\centering
\includegraphics[scale=0.75]{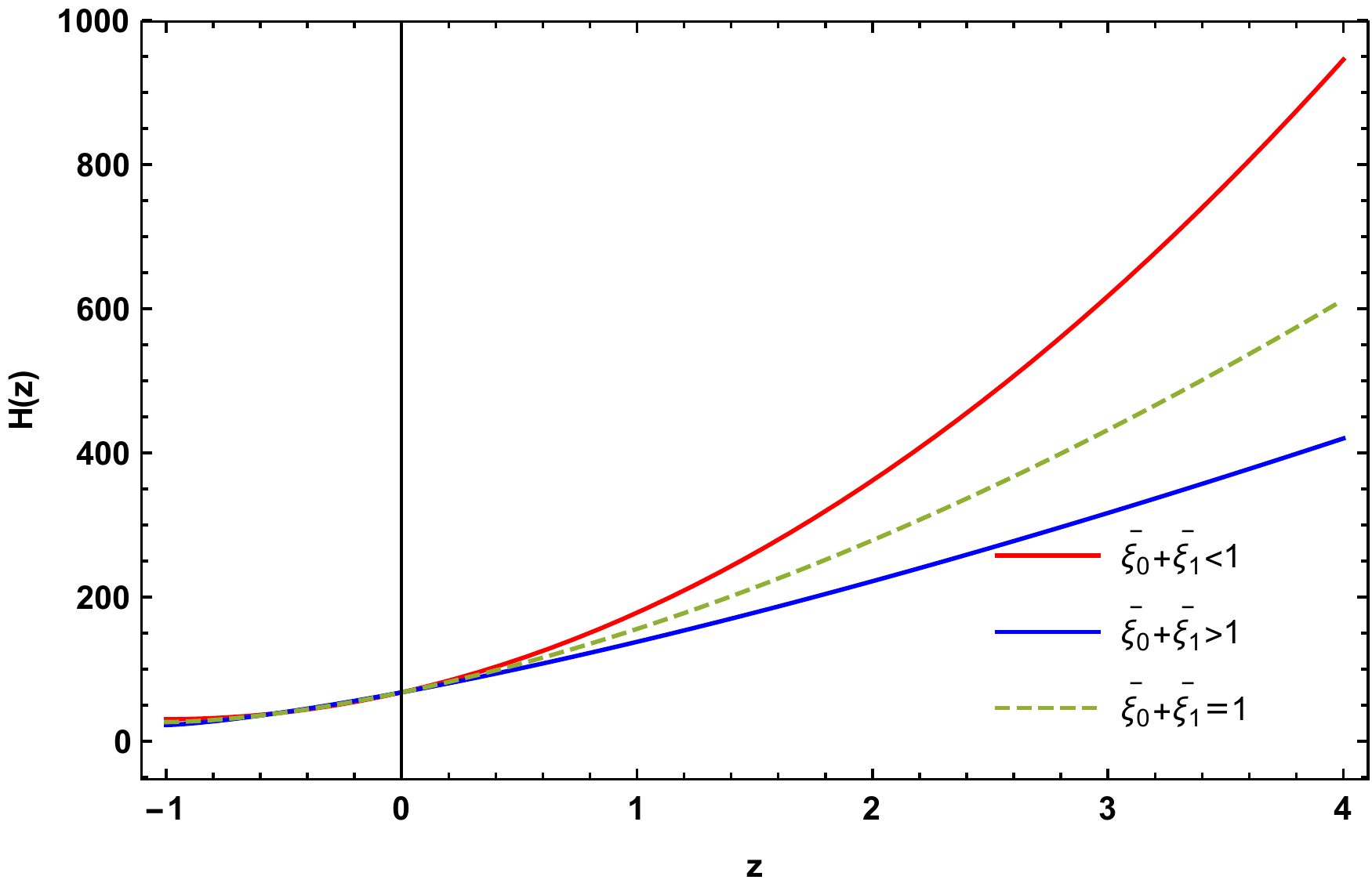}
\caption{Variation of the Hubble parameter with redshift $z$ for the first
limiting conditions $\bar{\protect\xi}_{0}>0, \bar{\protect\xi}_{12}<3, \bar{%
\protect\xi}_{2}<2, \bar{\protect\xi}_{0}+\bar{\protect\xi}_{12}<3$. Here we
took $\protect\alpha =-1$ i.e., $\bar{\protect\alpha}=1$. For Red, Blue and
Green plots the value of $(\bar{\protect\xi}_{0},\bar{\protect\xi}_{1},\bar{%
\protect\xi}_{2})$ are $(0.9,0.01,1),(0.45,0.65,1),(0.65,0.35,1)$
respectively.}
\label{f3}
\end{figure}

\begin{figure}[H]
\centering
\includegraphics[scale=0.75]{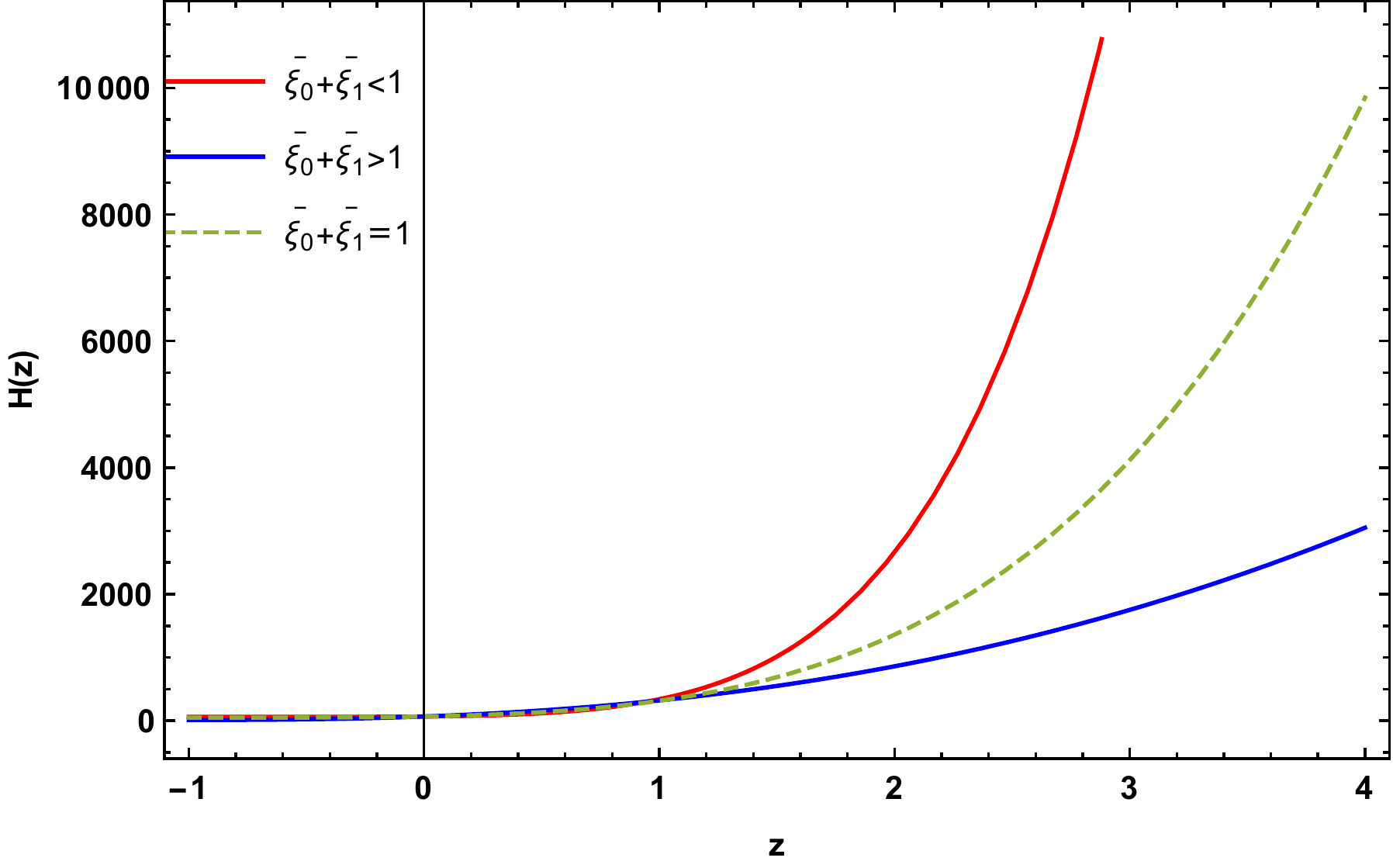}
\caption{Variation of the Hubble parameter with redshift $z$ for the second
limiting conditions $\bar{\protect\xi}_{0}<0, \bar{\protect\xi}_{12}>3, \bar{%
\protect\xi}_{2}>2, \bar{\protect\xi}_{0}+\bar{\protect\xi}_{12}>3$. Here we
took $\protect\alpha =-1$ i.e., $\bar{\protect\alpha}=1$. For Red, Blue and
Green plots the value of $(\bar{\protect\xi}_{0},\bar{\protect\xi}_{1},\bar{%
\protect\xi}_{2})$ are $(-0.5,1.45,2.1),(-0.5,2.5,3),(-0.5,1.5,2.17)$
respectively}
\label{f4}
\end{figure}

From the above figures of $q(z)\sim z$, we can see that, only first limiting
condition with $\bar{\xi}_{0}+\bar{\xi}_{1}>1$ (blue line in Fig. \ref{f1}
shows a phase transition from early deceleration to present acceleration and
second limiting condition is not be suitable to discuss the present
observational scenario. Also the first limiting condition with $\bar{\xi}%
_{0}+\bar{\xi}_{1}=1$ and $\bar{\xi}_{0}+\bar{\xi}_{1}<1$, which can be
inferred from the following plots of Hubble parameter $H(z)\sim z$ as shown
in Figs. \ref{f3} and \ref{f4}.


With a discussion of the geometrical behavior of our obtained model, we
shall now try to find out suitable numerical values of the model parameters
consistent with the present observations. We only consider the first
limiting condition for our further analysis.

\section{Best fit values of model parameters from observation}\label{sec6}

We have found an exact solution of the Einstein field equations with the
bulk viscous matter in $f(Q)$ gravity having four model parameters $\alpha $%
, $\xi _{0}$, $\xi _{1}$ and $\xi _{2}$. We must discuss the approximate
values of these model parameters describing well the present universe
through some observational datasets. To obtain the best fit values of these
model parameters, we have used mainly two datasets namely, the Hubble
datasets containing $57$ data points and the Pantheon datasets containing $%
1048$ datasets. Furthermore, we have discussed our results together with the
Baryon Acoustic oscillations (BAO) datasets. To constrain the model
parameters with the above discussed datasets, we have used the Python's
Scipy optimization technique. First, we have estimated the global minima for
the Hubble function in equation \eqref{3o}. For the numerical analysis, we
employ the Python's emcee library and consider a Gaussian prior with above
estimates as means and a fixed $\sigma =1.0$ as dispersion. The idea behind
the analysis is to check the parameter space in the neighborhood of the
local minima. More about the Hubble datasets, Pantheon datasets and BAO
datasets and methodology are discussed below in some detail and finally, the
results are discussed as 2-dimensional contour plots with $1-\sigma $ \& $%
2-\sigma $ errors.

\subsection{H(z) datasets}

We are familiar with the well known cosmological principle which assumes
that on the large scale, our universe is homogeneous and isotropic. This
principle is the backbone of modern cosmology. In the last few decades this
principle had been tested several times and have been supported by many
cosmological observations. In the study of observational cosmology, the
expansion scenario of the universe be directly investigated by the Hubble
parameter i.e. $H=\frac{\dot{a}}{a}$ where $\dot{a}$ represents derivative
of cosmic scale factor $a$ with respect to cosmic time $t$. The Hubble
parameter as a function of redshift can be expressed as $H(z)$ $=-\frac{1}{%
1+z}\frac{dz}{dt}$, where $dz$ is acquired from the spectroscopic surveys
and therefore a measurement of $dt$ furnishes the model independent value of
the Hubble parameter. In general, there are two well known methods that are
used to measure the value of the Hubble parameter values $H(z)$ at some
definite redshift. The first one is the extraction of $H(z)$ from
line-of-sight BAO data and another one is the differential age method \cite%
{H1}-\cite{H19}. In this manuscript, we have taken an updated set of $57$
data points. In this set of $57$ Hubble data points, $31$ points measured
via the method of differential age (DA) and remaining $26$ points through
BAO and other methods in the range of redshift given as $0.07\leqslant
z\leqslant 2.42$ \cite{sharov}. Furthermore, we have taken $H_{0}=69$ $%
Km/s/Mpc$ for our analysis . To find out the mean values of the model
parameters $\alpha $, $\xi _{0}$, $\xi _{1}$ and $\xi _{2}$ (which is
equivalent to the maximum likelihood analysis), we have taken the chi-square
function as,

\begin{equation}
\chi _{H}^{2}(\alpha ,\xi _{0},\xi _{1},\xi _{2})=\sum\limits_{i=1}^{57}%
\frac{[H_{th}(z_{i},\alpha ,\xi _{0},\xi _{1},\xi _{2})-H_{obs}(z_{i})]^{2}}{%
\sigma _{H(z_{i})}^{2}},  \label{chihz}
\end{equation}%
where the theoretical value of Hubble parameter is represented by $H_{th}$
and the observed value by $H_{obs}$ and $\sigma _{H(z_{i})}$ represents the
standard error in the observed value of $H$. The $57$ points of Hubble
parameter values $H(z)$ with errors $\sigma _{H}$ from differential age ($31$
points) method and BAO and other ($26$ points) methods are tabulated in
Table-1 with references.

\begin{center}
\begin{tabular}{|c|c|c|c|c|c|c|c|}\hline
\multicolumn{8}{|c|}{Table-1: 57 points of $H(z)$ datasets} \\ \hline
\multicolumn{8}{|c|}{31 points from DA method}  \\ \hline
$z$ & $H(z)$ & $\sigma _{H}$ & Ref. & $z$ & $H(z)$ & $\sigma _{H}$ & Ref. \\ \hline
$0.070$ & $69$ & $19.6$ & \cite{H1} & $0.4783$ & $80$ & $99$ & \cite{H5} \\ \hline
$0.90$ & $69$ & $12$ & \cite{H2} & $0.480$ & $97$ & $62$ & \cite{H1} \\ \hline
$0.120$ & $68.6$ & $26.2$ & \cite{H1} & $0.593$ & $104$ & $13$ & \cite{H3} \\ \hline
$0.170$ & $83$ & $8$ & \cite{H2} & $0.6797$ & $92$ & $8$ & \cite{H3} \\ \hline
$0.1791$ & $75$ & $4$ & \cite{H3} & $0.7812$ & $105$ & $12$ & \cite{H3} \\ \hline
$0.1993$ & $75$ & $5$ & \cite{H3} & $0.8754$ & $125$ & $17$ & \cite{H3} \\ \hline
$0.200$ & $72.9$ & $29.6$ & \cite{H4} & $0.880$ & $90$ & $40$ & \cite{H1} \\ \hline
$0.270$ & $77$ & $14$ & \cite{H2} & $0.900$ & $117$ & $23$ & \cite{H2} \\ \hline 
$0.280$ & $88.8$ & $36.6$ & \cite{H4} & $1.037$ & $154$ & $20$ & \cite{H3} \\ \hline 
$0.3519$ & $83$ & $14$ & \cite{H3} & $1.300$ & $168$ & $17$ & \cite{H2} \\ \hline 
$0.3802$ & $83$ & $13.5$ & \cite{H5} & $1.363$ & $160$ & $33.6$ & \cite{H7} \\ \hline 
$0.400$ & $95$ & $17$ & \cite{H2} & $1.430$ & $177$ & $18$ & \cite{H2} \\ \hline 
$0.4004$ & $77$ & $10.2$ & \cite{H5} & $1.530$ & $140$ & $14$ & \cite{H2} \\ \hline
$0.4247$ & $87.1$ & $11.2$ & \cite{H5} & $1.750$ & $202$ & $40$ & \cite{H2} \\ \hline
$0.4497$ & $92.8$ & $12.9$ & \cite{H5} & $1.965$ & $186.5$ & $50.4$ & \cite{H7}  \\ \hline
$0.470$ & $89$ & $34$ & \cite{H6} &  &  &  &   \\ \hline
\multicolumn{8}{|c|}{26 points from BAO \& other method} \\ \hline
$z$ & $H(z)$ & $\sigma _{H}$ & Ref. & $z$ & $H(z)$ & $\sigma _{H}$ & Ref. \\ \hline
$0.24$ & $79.69$ & $2.99$ & \cite{H8} & $0.52$ & $94.35$ & $2.64$ & \cite{H10} \\ \hline
$0.30$& $81.7$ & $6.22$ & \cite{H9} & $0.56$ & $93.34$ & $2.3$ & \cite{H10} \\ \hline
$0.31$ & $78.18$ & $4.74$ & \cite{H10} & $0.57$ & $87.6$ & $7.8$ & \cite{H14} \\ \hline
$0.34$ & $83.8$ & $3.66$ & \cite{H8} & $0.57$ & $96.8$ & $3.4$ & \cite{H15} \\ \hline
$0.35$ & $82.7$ & $9.1$ & \cite{H11} & $0.59$ & $98.48$ & $3.18$ & \cite{H10} \\ \hline
$0.36$ & $79.94$ & $3.38$ & \cite{H10} & $0.60$ & $87.9$ & $6.1$ & \cite{H13} \\ \hline
$0.38$ & $81.5$ & $1.9$ & \cite{H12} & $0.61$ & $97.3$ & $2.1$ & \cite{H12} \\ \hline
$ 0.40$ & $82.04$ & $2.03$ & \cite{H10} & $0.64$ & $98.82$ & $2.98$ & \cite{H10}  \\ \hline
$0.43$ & $86.45$ & $3.97$ & \cite{H8} & $0.73$ & $97.3$ & $7.0$ & \cite{H13} \\ \hline
$0.44$ & $82.6$ & $7.8$ & \cite{H13} & $2.30$ & $224$ & $8.6$ & \cite{H16} \\ \hline
$0.44$ & $84.81$ & $1.83$ & \cite{H10} & $2.33$ & $224$ & $8$ & \cite{H17} \\ \hline
$0.48$ & $87.79$ & $2.03$ & \cite{H10} & $2.34$ & $222$ & $8.5$ & \cite{H18} \\ \hline
$0.51$ & $90.4$ & $1.9$ & \cite{H12} & $2.36$ & $226$ & $9.3$ & \cite{H19} \\ \hline
\end{tabular}
\end{center}

\begin{figure}[H]
\centering
\includegraphics[scale=0.95]{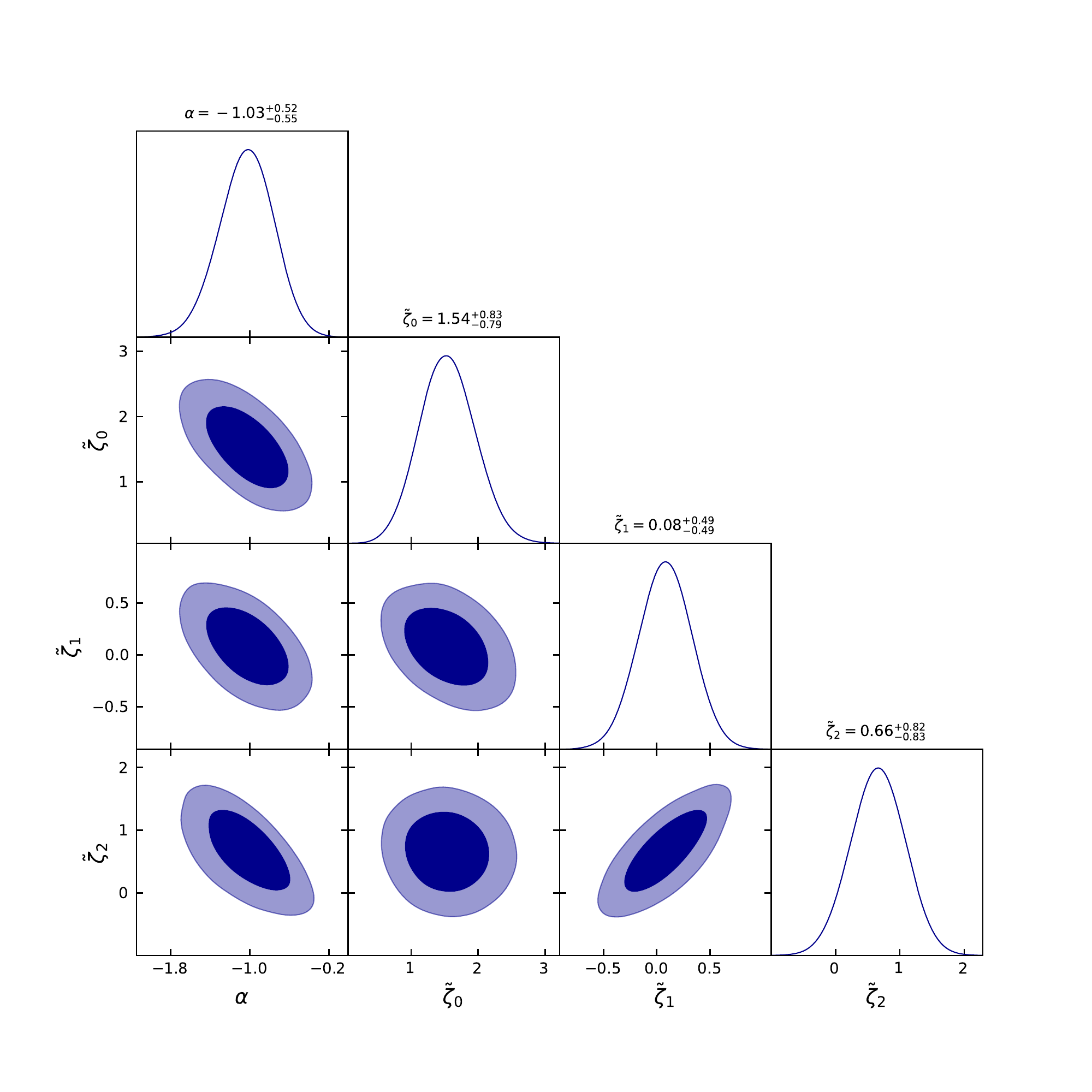}  
\caption{The plot shows the 2-d contour plots of the model parameters with $1-\protect%
\sigma $ and $2-\protect\sigma $ errors and also shows the best fit values of the model parameters $%
\protect\alpha $, $\protect\xi _{0}$, $\protect\xi _{1}$ and $\protect\xi %
_{2}$ obtained from the $57$ points of Hubble datasets.}
\label{fig:Hubble}
\end{figure}

\begin{figure}[H]
\centering
\includegraphics[scale=0.65]{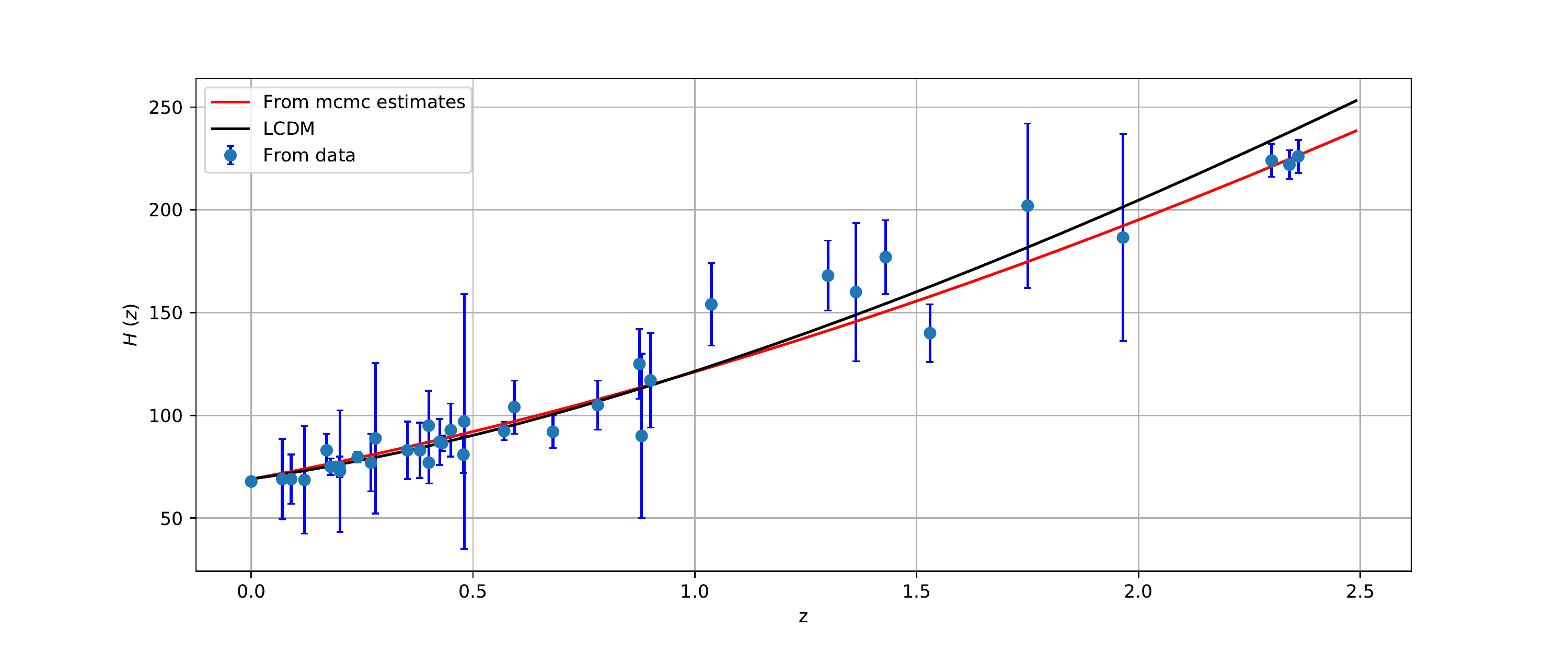}  
\caption{The plot shows the plot of Hubble function $H(z)$ vs.
redshift $z$ for our model shown in red line which shows nice fit to the $57$
points of the Hubble datasets shown in dots with it's error bars and also compared to the $\Lambda$CDM model shown in black solid line with $\Omega_{m0}=0.3$ \& $\Omega_{\Lambda 0}=0.7$.}
\label{fig:final-hz}
\end{figure}

Using the above datasets, we have estimated the best fit
values of the model parameters $\alpha $, $\xi _{0}$, $\xi _{1}$ and $\xi
_{2}$ as and is shown in the following plot \ref{fig:Hubble} as 2-d contour
sub-plots with $1-\sigma $ \& $2-\sigma $ errors. The best fit values are
obtained as $\alpha =-1.03_{-0.55}^{+0.52}$, $\xi _{0}=1.54_{-0.79}^{+0.83}$%
, $\xi _{1}=0.08_{-0.49}^{+0.49}$ and $\xi _{2}=0.66_{-0.83}^{+0.82}$ with
the $57$ points of Hubble datasets as given in Table-1. 
Also, we have shown
the error bar plot for the discussed Hubble datasets and is shown in the
following plot \ref{fig:final-hz} together with our obtained model compared
with the $\Lambda $CDM model (with $\Omega _{m0}=0.3$ and $\Omega _{\Lambda
0}=0.7$). The plot shows nice fit of our model to the observational Hubble
datasets.

\newpage
\subsection{Pantheon datasets}

Initially, the observational studies on supernovae of the golden sample of $%
50$ points of type \textit{Ia} suggested that our universe is in an
accelerating phase of expansion. After the result, the studies on more and
more samples of supernovae datasets increased during the past two decades.
Recently, the latest sample of supernovae of type \textit{Ia} datasets are
released containing $1048$ data points. In this article, we have used this
set of datasets known as Pantheon datasets \cite{DM} with $1048$ samples of
spectroscopically confirmed SNe \textit{Ia} covering the range in the
redshift range $0.01<z<2.26$. In the redshift range $0<z_{i}\leq 1.41$,
these data points gives the estimation of the distance moduli $\mu _{i}=\mu
_{i}^{obs}$. Here, we fit our model parameters of the obtained model,
comparing the theoretical $\mu _{i}^{th}$ value and the observed $\mu
_{i}^{obs}$ value of the distance modulus. The distance moduli which are the
logarithms given as $\mu _{i}^{th}=\mu (D_{L})=m-M=5\log _{10}(D_{L})+\mu
_{0}$, where $m$ and $M$ represents apparent and absolute magnitudes and $%
\mu _{0}=5\log \left( H_{0}^{-1}/Mpc\right) +25$ is the marginalized
nuisance parameter. The luminosity distance is taken to be, 
\begin{eqnarray*}
D_{l}(z) &=&\frac{c(1+z)}{H_{0}}S_{k}\left( H_{0}\int_{0}^{z}\frac{1}{%
H(z^{\ast })}dz^{\ast }\right) , \\
\text{where }S_{k}(x) &=&\left\{ 
\begin{array}{c}
\sinh (x\sqrt{\Omega _{k}})/\Omega _{k}\text{, }\Omega _{k}>0 \\ 
x\text{, \ \ \ \ \ \ \ \ \ \ \ \ \ \ \ \ \ \ \ \ \ \ \ }\Omega _{k}=0 \\ 
\sin x\sqrt{\left\vert \Omega _{k}\right\vert })/\left\vert \Omega
_{k}\right\vert \text{, }\Omega _{k}<0%
\end{array}%
\right. .
\end{eqnarray*}%
Here, $\Omega _{k}=0$ (flat space-time). We have calculated distance $%
D_{L}(z)$ and corresponding chi square function that measures difference
between predictions of our model and the SN \textit{Ia} observational data.
The $\chi _{SN}^{2}$ function for the Pantheon datasets is taken to be,

\begin{equation}
\chi _{SN}^{2}(\mu _{0},\alpha ,\xi _{0},\xi _{1},\xi
_{2})=\sum\limits_{i=1}^{1048}\frac{[\mu ^{th}(\mu _{0},z_{i},\alpha ,\xi
_{0},\xi _{1},\xi _{2})-\mu ^{obs}(z_{i})]^{2}}{\sigma _{\mu (z_{i})}^{2}},
\label{chisn}
\end{equation}%
$\sigma _{\mu (z_{i})}^{2}$ is the standard error in the observed value.
After marginalizing $\mu _{0}$, the chi square function is written as,

$\qquad \qquad \qquad \qquad \qquad \qquad \qquad \qquad \chi
_{SN}^{2}(\alpha ,\xi _{0},\xi _{1},\xi _{2})=A(\alpha ,\xi _{0},\xi
_{1},\xi _{2})-[B(\alpha ,\xi _{0},\xi _{1},\xi _{2})]^{2}/C(\alpha ,\xi
_{0},\xi _{1},\xi _{2})$

where

$\qquad \qquad \qquad \qquad \qquad \qquad \qquad \qquad A(\alpha ,\xi
_{0},\xi _{1},\xi _{2})=\sum\limits_{i=1}^{1048}\frac{[\mu ^{th}(\mu
_{0}=0,z_{i},\alpha ,\xi _{0},\xi _{1},\xi _{2})-\mu ^{obs}(z_{i})]^{2}}{%
\sigma _{\mu (z_{i})}^{2}},$

$\qquad \qquad \qquad \qquad \qquad \qquad \qquad \qquad B(\alpha ,\xi
_{0},\xi _{1},\xi _{2})=\sum\limits_{i=1}^{1048}\frac{[\mu ^{th}(\mu
_{0}=0,z_{i},\alpha ,\xi _{0},\xi _{1},\xi _{2})-\mu ^{obs}(z_{i})]^{2}}{%
\sigma _{\mu (z_{i})}^{2}},$

$\qquad \qquad \qquad \qquad \qquad \qquad \qquad \qquad C(\alpha ,\xi
_{0},\xi _{1},\xi _{2})=\sum\limits_{i=1}^{1048}\frac{1}{\sigma _{\mu
(z_{i})}^{2}}$.

\begin{figure}[H]
\centering
\includegraphics[scale=0.95]{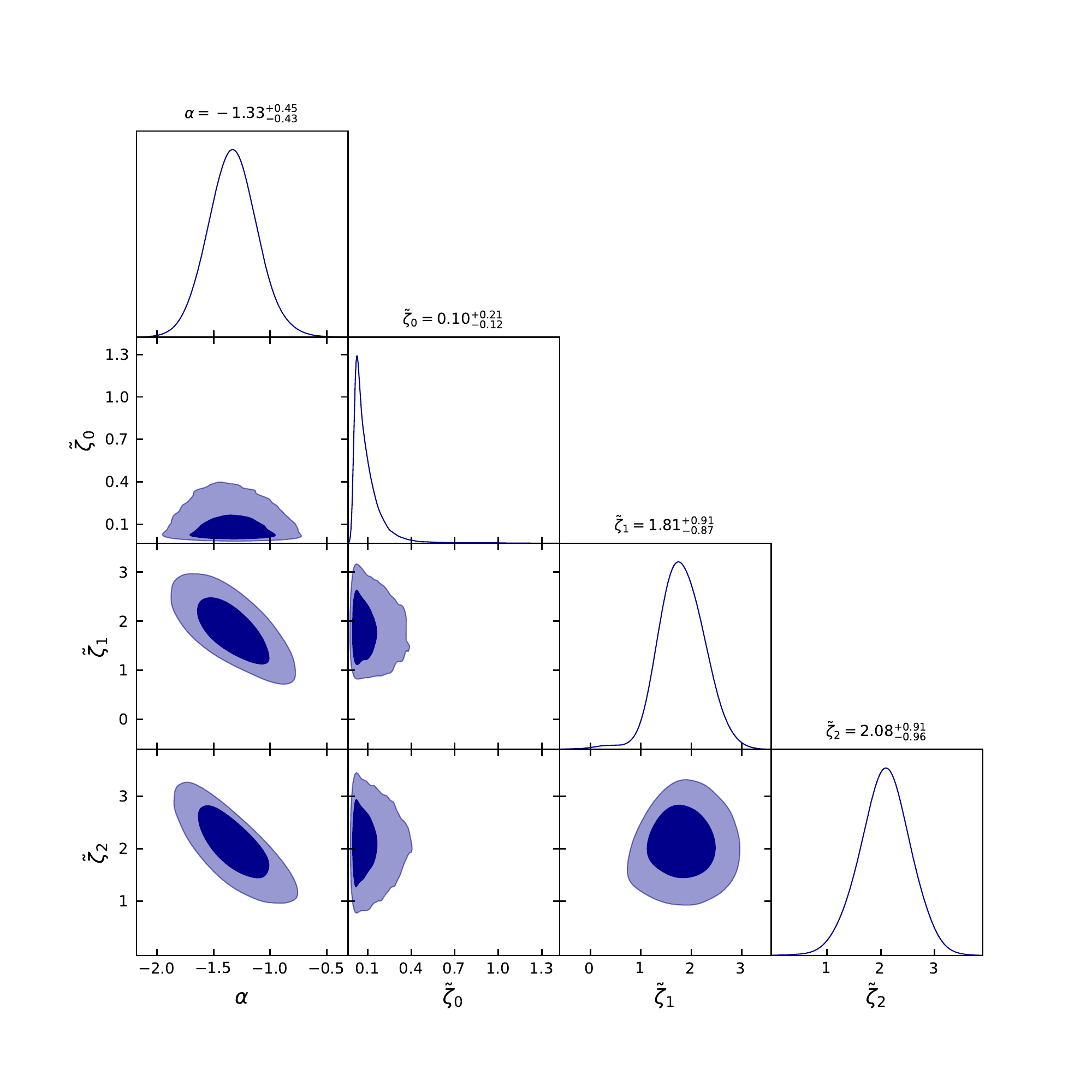}  
\caption{The plot shows the best fit values of the model parameters $%
\protect\alpha $, $\protect\xi _{0}$, $\protect\xi _{1}$ and $\protect\xi %
_{2}$ obtained w.r.t to the $1048$ points of Pantheon datasets at $1-\protect%
\sigma $ and $2-\protect\sigma $ confidence level.}
\label{fig:Pantheon}
\end{figure}

Using the above Pantheon datasets, we have estimated the best fit values of
the model parameters $\alpha $, $\xi _{0}$, $\xi _{1}$ and $\xi _{2}$ and is
shown in the following plot \ref{fig:Pantheon} as 2-d contour sub-plots with 
$1-\sigma $ \& $2-\sigma $ errors. The best fit values are obtained as $%
\alpha =-1.33_{-0.43}^{+0.45}$, $\xi _{0}=0.10_{-0.12}^{+0.21}$, $\xi
_{1}=1.81_{-0.87}^{+0.91}$ and $\xi _{2}=2.08_{-0.96}^{+0.91}$ with $1048$
points of Pantheon datasets. Also, we have shown the error bar plot for the
discussed Pantheon datasets and is shown in the following plot \ref%
{fig:final-muz} together with our obtained model compared with the $\Lambda $%
CDM model (with $\Omega _{m0}=0.3$ and $\Omega _{\Lambda 0}=0.7$). The plot
shows nice fit of our model to the observational Pantheon datasets.

\begin{figure}[H]
\centering
\includegraphics[scale=0.65]{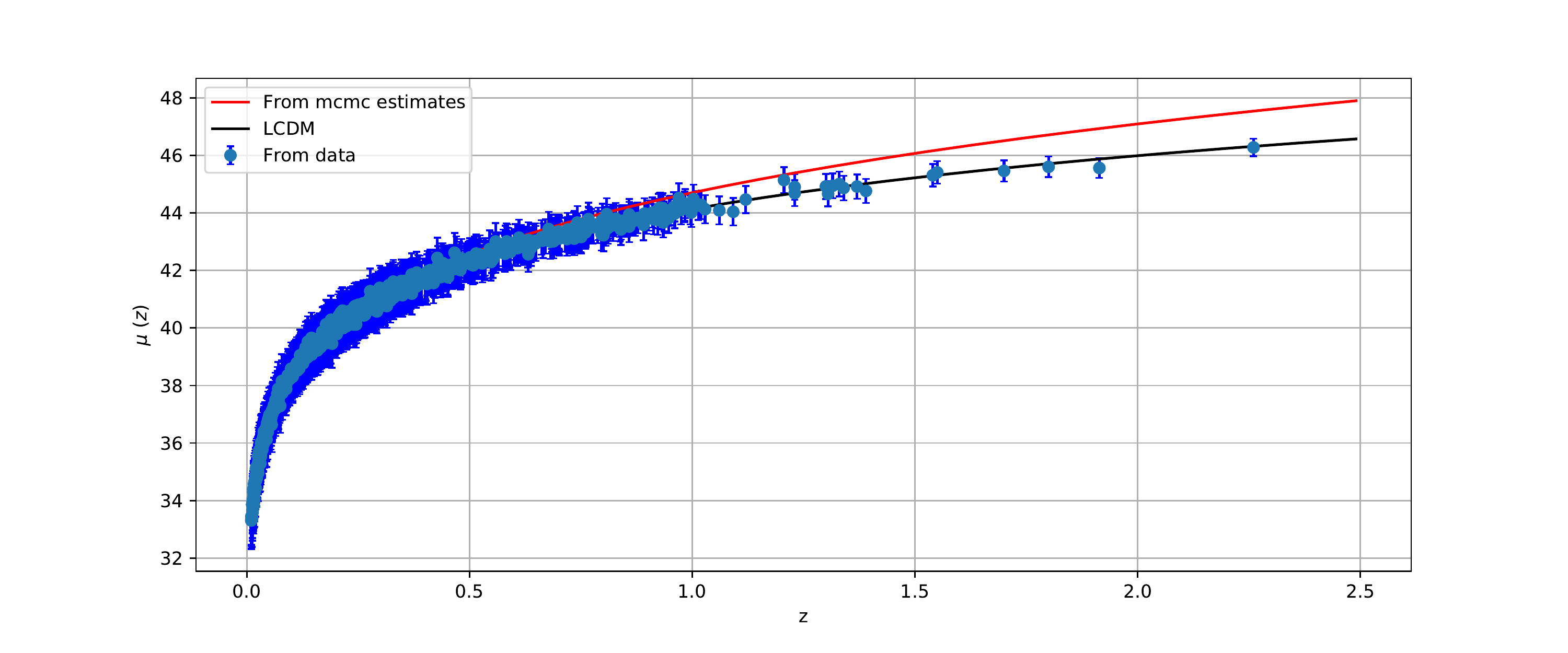}  
\caption{{}{}The plot shows the plot of distance modulus $\protect\mu (z)$
vs. redshift $z$ for our model shown in red line which shows nice fit to the 
$1048$ points of the Pantheon datasets shown in dots with it's error bars.}
\label{fig:final-muz}
\end{figure}

\subsection{BAO datasets}

In the study of the Early Universe baryons, photons and dark
matter come into picture and act as a single fluid coupled tightly through
the Thompson scattering) but do not collapse under gravity and oscillate due
to the large pressure of photons. Baryonic acoustic oscillations is an
analysis that discusses these oscillations in the early Universe. The
characteristic scale of BAO is governed by the sound horizon $r_{s}$ at the photon decoupling epoch $z_{\ast }$ and is given by the
following relation, 
\begin{equation*}
r_{s}(z_{\ast })=\frac{c}{\sqrt{3}}\int_{0}^{\frac{1}{1+z_{\ast }}}\frac{da}{%
a^{2}H(a)\sqrt{1+(3\Omega _{0b}/4\Omega _{0\gamma })a}},
\end{equation*}%
where $\Omega _{0b}$ and $\Omega _{0\gamma }$
are respectively referred to baryon density and photon density at present
time.

The angular diameter distance $D_{A}$ and the Hubble
expansion rate $H$ as a function of $z$ are derived
using the BAO sound horizon scale. If the measured angular separation of the
BAO feature is denoted by $\triangle \theta $ in the 2 point
correlation function of the galaxy distribution on the sky and if the
measured redshift separation of the BAO feature is denoted by $\triangle z$%
 in same the 2 point correlation function along the line of sight
then we have the relation, $\triangle \theta =\frac{r_{s}}{d_{A}(z)}$%
 where $d_{A}(z)=\int_{0}^{z}\frac{dz^{\prime }}{H(z^{\prime })}$%
 and $\triangle z=H(z)r_{s}$. Here, in this work, a simple
BAO datasets of six points for $d_{A}(z_{\ast })/D_{V}(z_{BAO})$
is considered from the references \cite{BAO1, BAO2, BAO3, BAO4, BAO5, BAO6},
where the photon decoupling redshift is taken as $z_{\ast }\approx 1091$%
 and $d_{A}(z)$ is the co-moving angular diameter
distance together with the dilation scale $D_{V}(z)=\left(
d_{A}(z)^{2}z/H(z)\right) ^{1/3}$. The following table-2 shows the
six points of the BAO datasets,

\begin{center}
\begin{tabular}{|c|c|c|c|c|c|c|}
\hline
\multicolumn{7}{|c|}{Table-2: Values of $d_{A}(z_{\ast })/D_{V}(z_{BAO})$
for distinct values of $z_{BAO}$} \\ \hline
$z_{BAO}$ & $0.106$ & $0.2$ & $0.35$ & $0.44$ & $0.6$ & $0.73$ \\ \hline
$\frac{d_{A}(z_{\ast })}{D_{V}(z_{BAO})}$ & $30.95\pm 1.46$ & $17.55\pm 0.60$
& $10.11\pm 0.37$ & $8.44\pm 0.67$ & $6.69\pm 0.33$ & $5.45\pm 0.31$ \\ 
\hline
\end{tabular}%
\textbf{.}
\end{center}

\qquad Also, the chi square function for BAO is given by \cite{BAO6}%
, 
\begin{equation}
\chi _{BAO}^{2}=X^{T}C^{-1}X\,,  \label{chibao}
\end{equation}%
\textbf{where }%
\begin{equation*}
X=\left( 
\begin{array}{c}
\frac{d_{A}(z_{\star })}{D_{V}(0.106)}-30.95 \\ 
\frac{d_{A}(z_{\star })}{D_{V}(0.2)}-17.55 \\ 
\frac{d_{A}(z_{\star })}{D_{V}(0.35)}-10.11 \\ 
\frac{d_{A}(z_{\star })}{D_{V}(0.44)}-8.44 \\ 
\frac{d_{A}(z_{\star })}{D_{V}(0.6)}-6.69 \\ 
\frac{d_{A}(z_{\star })}{D_{V}(0.73)}-5.45%
\end{array}%
\right) \,,
\end{equation*}
and the inverse covariance matrix $C^{-1}$ is defined in 
\cite{BAO6}. 

\begin{equation*}
C^{-1}=\left( 
\begin{array}{cccccc}
0.48435 & -0.101383 & -0.164945 & -0.0305703 & -0.097874 & -0.106738 \\ 
-0.101383 & 3.2882 & -2.45497 & -0.0787898 & -0.252254 & -0.2751 \\ 
-0.164945 & -2.454987 & 9.55916 & -0.128187 & -0.410404 & -0.447574 \\ 
-0.0305703 & -0.0787898 & -0.128187 & 2.78728 & -2.75632 & 1.16437 \\ 
-0.097874 & -0.252254 & -0.410404 & -2.75632 & 14.9245 & -7.32441 \\ 
-0.106738 & -0.2751 & -0.447574 & 1.16437 & -7.32441 & 14.5022%
\end{array}%
\right) \,.
\end{equation*}

\begin{figure}[H]
\centering
\includegraphics[scale=0.9]{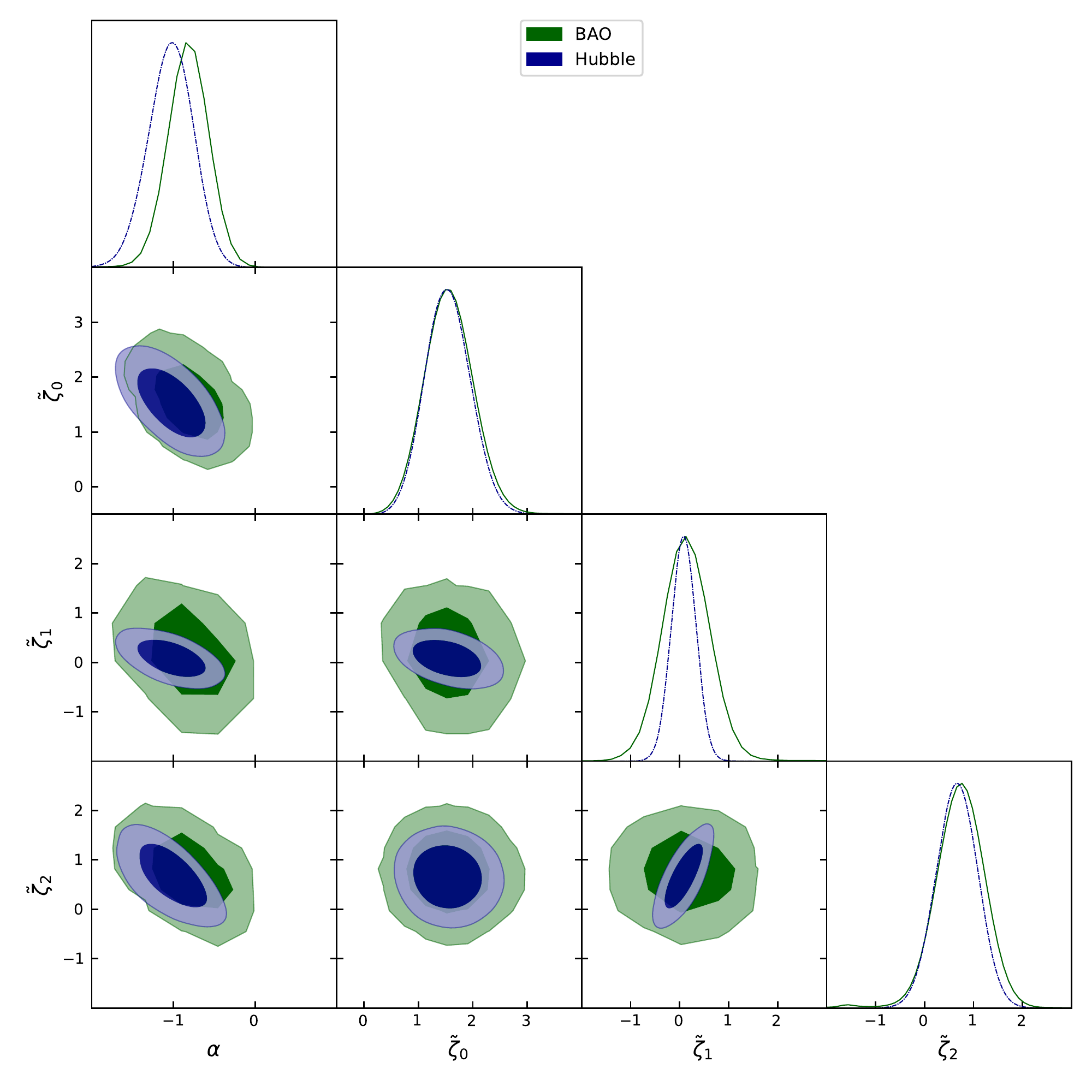}  
\caption{The plot shows the 2-d contour plots of the model parameters with $1-\protect%
\sigma $ and $2-\protect\sigma $ errors and also shows the best fit values of the model parameters $%
\protect\alpha $, $\protect\xi _{0}$, $\protect\xi _{1}$ and $\protect\xi %
_{2}$ obtained from the $57$ points of Hubble datasets together with six points of BAO datasets.}
\label{fig:Hubble_BAO}
\end{figure}

\begin{figure}[H]
\centering
\includegraphics[scale=0.9]{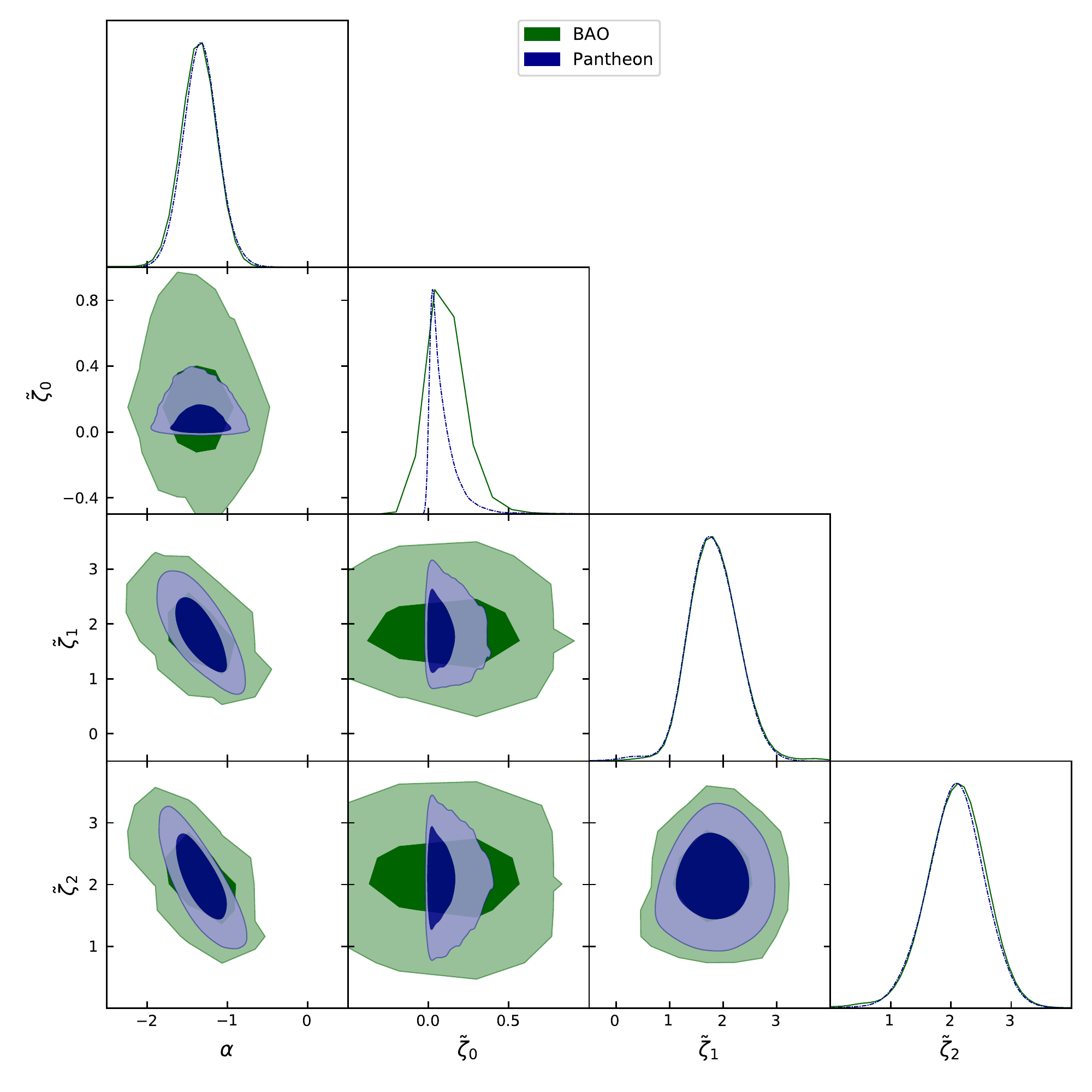}  
\caption{The plot shows the best fit values of the model parameters $%
\protect\alpha $, $\protect\xi _{0}$, $\protect\xi _{1}$ and $\protect\xi %
_{2}$ obtained w.r.t to the $1048$ points of Pantheon datasets together with six points of BAO datasets at $1-\protect%
\sigma $ and $2-\protect\sigma $ confidence level.}
\label{fig:Pantheon_BAO}
\end{figure}

Including these six data points of the BAO datasets with the Hubble
datasets and combine the above results of Hubble constrained values, we
obtain the values of the model parameters as, $\alpha
=-1.06_{-0.82}^{+0.34} $, $\xi _{0}=2.25_{-1.7}^{+0.37}$, $%
\xi _{1}=-0.08_{-0.96}^{+1.1}$\textbf{\ and }$\xi _{2}=0.7_{-1.1}^{+1.1}$%
. Similarly, including these six data points of the BAO datasets
with the Pantheon datasets and combine the above results of Panteon
constrained values, we obtain the values of the model parameters as, $%
\alpha =-1.65_{-0.25}^{+0.85}$, $\xi _{0}=0.86_{-1.1}^{+0.29}$%
, $\xi _{1}=-1.90_{-0.99}^{+0.97}$ and $\xi
_{2}=1.93_{-0.91}^{+1.2}$. The combined results are shown as 2-d
contour sub-plots with $1-\sigma $ \& $2-\sigma $ errors
in the following plots \ref{fig:Hubble_BAO} and \ref{fig:Pantheon_BAO}.

\section{Statefinder Diagnostic}\label{sec7}

It is well-known that the deceleration parameter $q$ and the Hubble
parameter $H$ are the oldest geometric variable. During the last few
decades, plenty of DE (Dark Energy) models have been proposed and the
remarkable growth in the precision of observational data both motivate us to
go beyond these two parameters. In this direction, a new pair of geometrical
quantities have been proposed by V. Sahni et al. \cite{V.Sahni} called
statefinder diagnostic parameters $(r,s) $. The state finder parameters
investigate the expansion dynamics through second and third derivatives of
the cosmic scale factor which is a natural succeeding step beyond the
parameters $H $ and $q $. These parameters are defined as follows

\begin{equation}
r=\frac{\dddot{a}}{aH^{3}}
\end{equation}%
and 
\begin{equation}
s=\frac{r-1}{3(q-\frac{1}{2})}\text{.}
\end{equation}

\begin{figure}[H]
\centering
\includegraphics[scale=0.55]{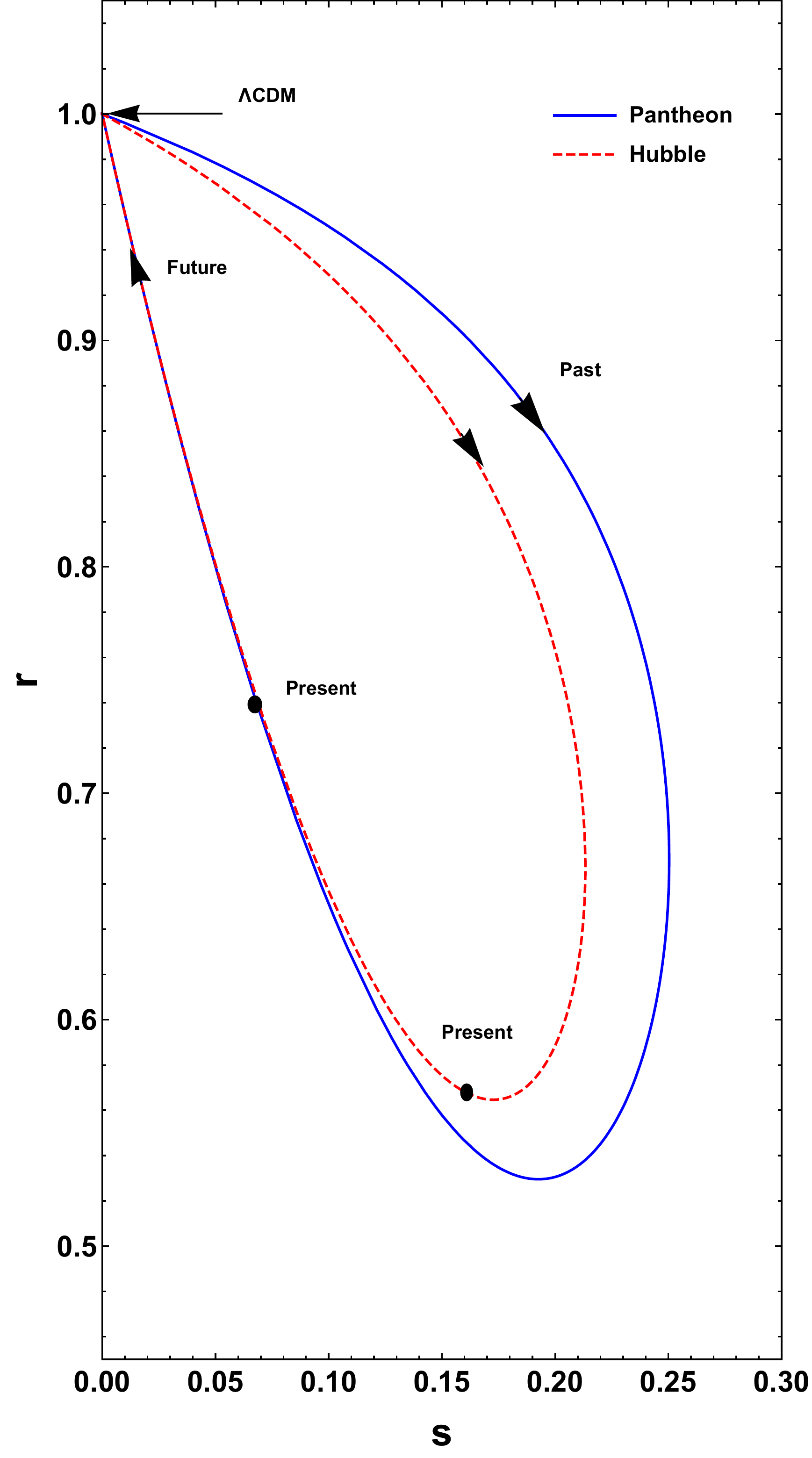}
\caption{The evolution trajectories of the given model in $s-r$ plane
corresponding to the values of model parameters constrained by the Hubble
and Pantheon datasets. }
\label{r-s}
\end{figure}

\begin{figure}[H]
\centering
\includegraphics[scale=0.85]{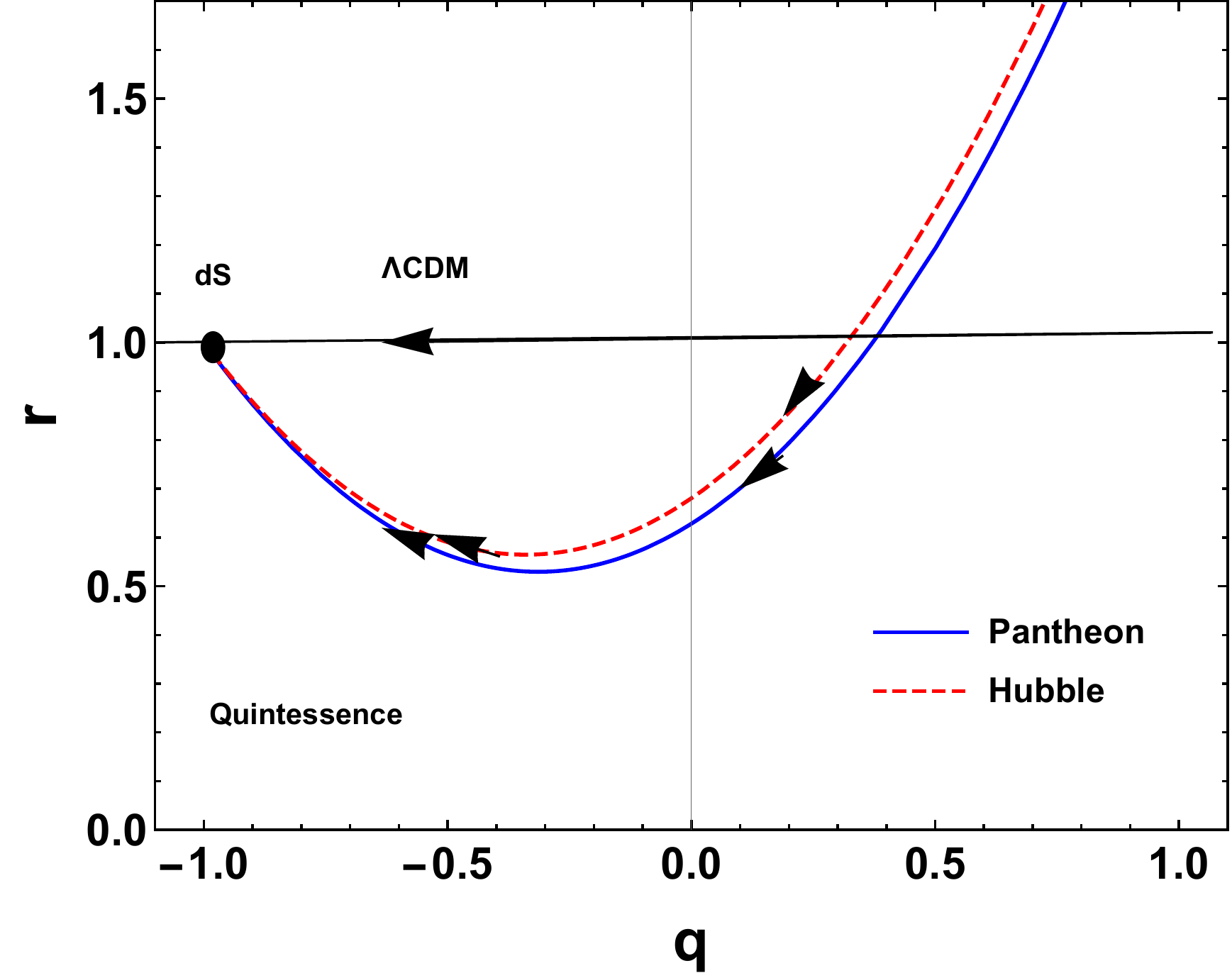}
\caption{The evolution trajectories of the given model in $q-r$ plane
corresponding to the values of model parameters constrained by the Hubble
and Pantheon datasets.}
\label{r-q}
\end{figure}

\begin{figure}[H]
\centering
\includegraphics[scale=0.75]{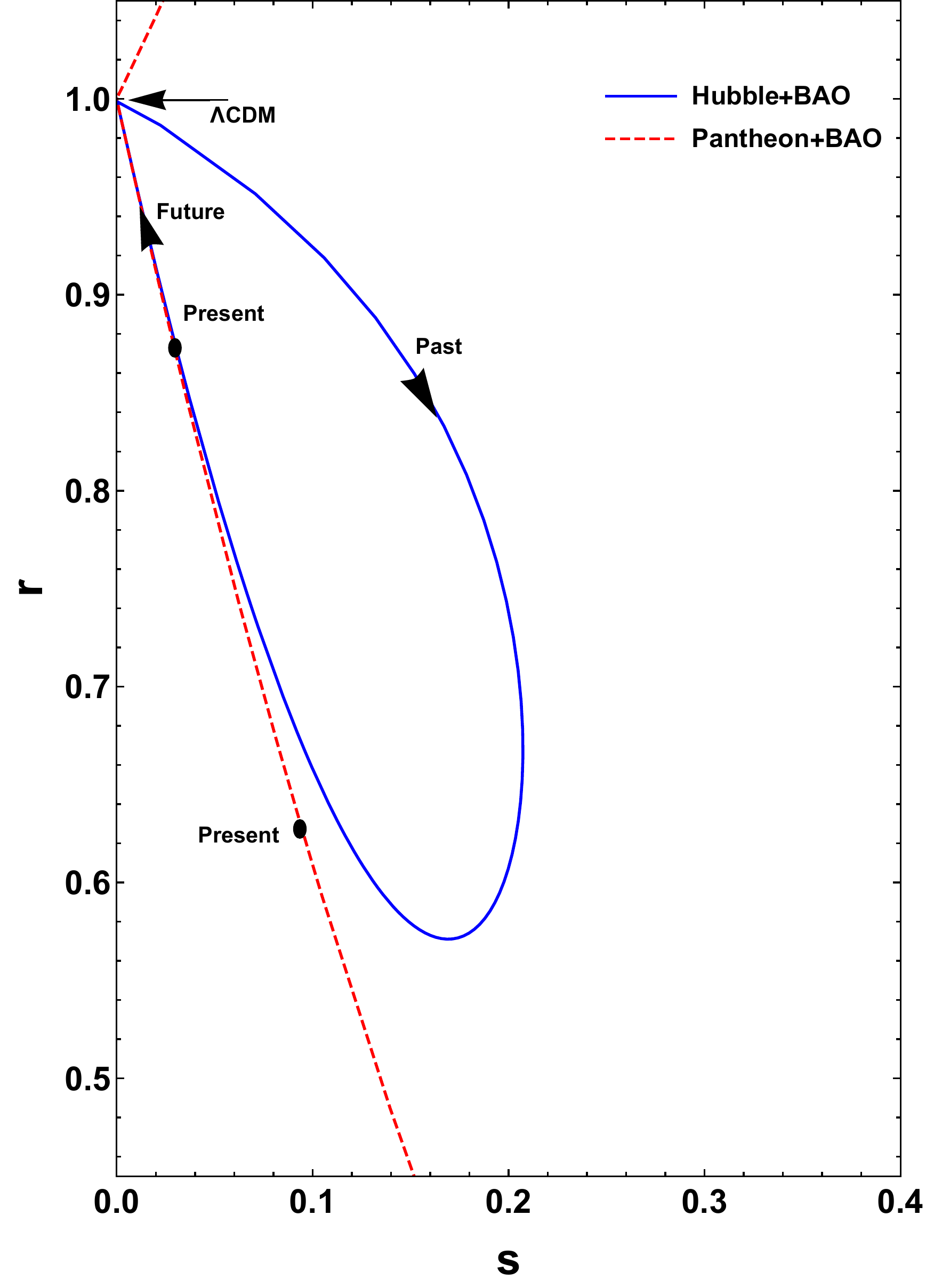}
\caption{The evolution trajectories of the given model in $s-r$ plane
corresponding to the values of model parameters constrained by the Hubble
and Pantheon datasets together with BAO datasets. }
\label{r-s-BAO}
\end{figure}

\begin{figure}[H]
\centering
\includegraphics[scale=0.6]{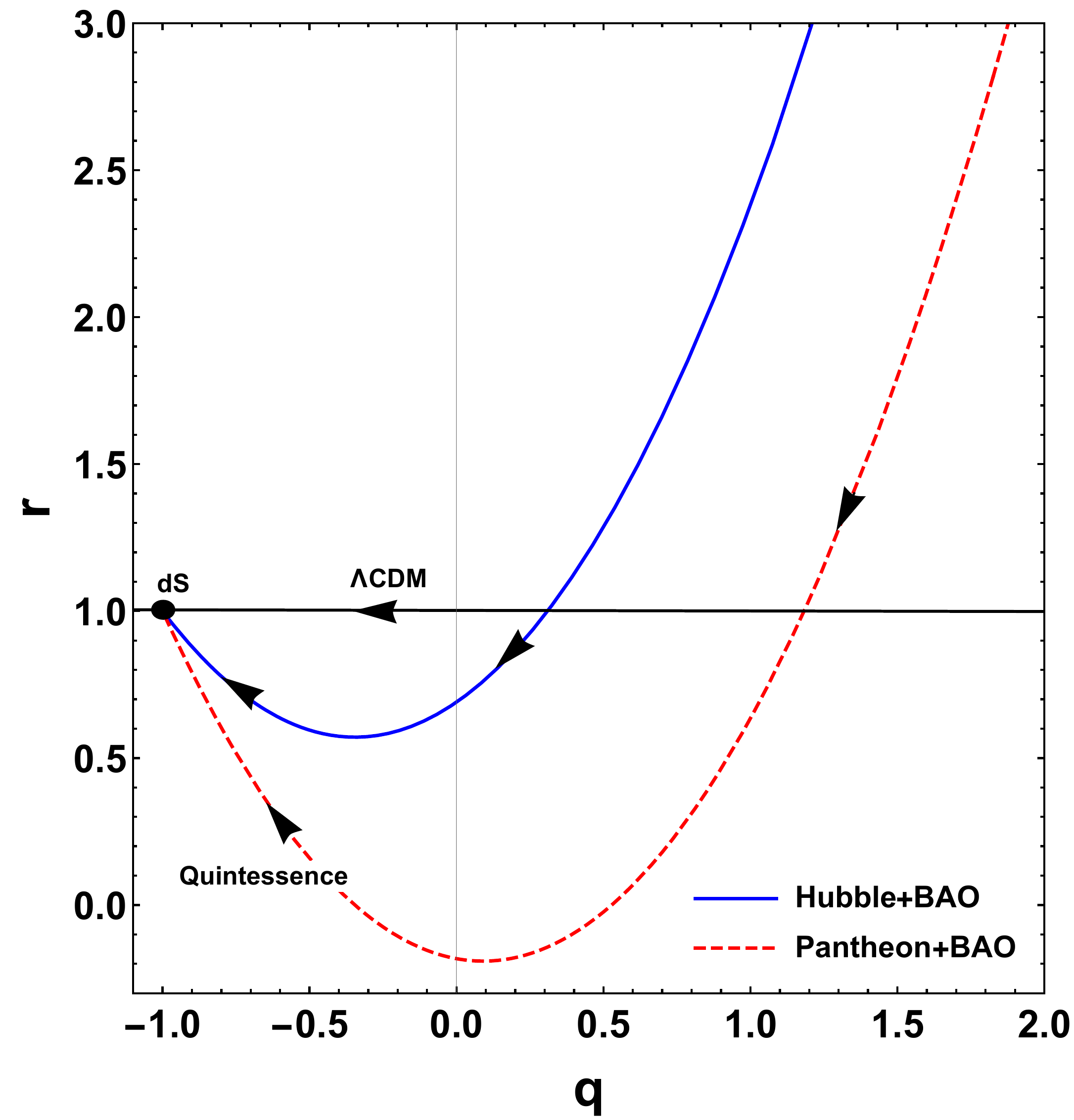}
\caption{The evolution trajectories of the given model in $q-r$ plane
corresponding to the values of model parameters constrained by the Hubble
and Pantheon datasets together with BAO datasets.}
\label{r-q-BAO}
\end{figure}

The fixed point $(s,r)=(0,1)$ in the $s-r$ diagram \ref{r-s} shows the
spatially flat $\Lambda $CDM model and $(q,r)=(-1,1)$ shows the de Sitter
point in Fig. \ref{r-q}. We plot the $s-r$ and $q-r$ diagram for the values
of $\alpha $, $\xi _{0}$, $\xi _{1}$ and $\xi _{2}$ constrained by the
Hubble and the Pantheon data sets. Corresponding to the Hubble and the
Pantheon datasets, the present values of $(s,r)$ parameter are $%
(0.159,0.568) $ and $(0.065,0.748)$ respectively.Furthermore, we
have also plotted the $s-r$ and $q-r$ diagrams for the
other set of values of the model parameters $\alpha $, $\xi _{0}$%
, $\xi _{1}$ and $\xi _{2}$ as obtained by the
combined results of BAO datasets with the Hubble and the Pantheon datasets
and are shown in the following plots \ref{r-s-BAO} \& \ref{r-q-BAO}. Corresponding to the combined results of BAO datasets with Hubble and the
Pantheon datasets, the present values of $(s,r)$ parameter are $%
(0.030,0.872) $ and $(0.096,0.621)$ respectively.

The statefinder diagnostic can differentiate the variety of dark energy
models like quintessence, the Chaplygin gas, braneworld models, etc. See the
references \cite{U.Alam,Gorini,Zim}. The departure of our bulk viscous model
from this fixed point establishes the distance of the given model from $%
\Lambda $CDM model. In present epoch the given model lie in Quintessence
region $(s>0,r<1)$. We can observe that the trajectories of $s-r$ diagram
will pass through the $\Lambda $CDM fixed point in the future. Thus the
statefinder diagnostic successfully shows that the given model is different
from other models of dark energy.

\section{Conclusions}\label{sec8}

In this article, we analyzed the evolution of FLRW universe dominated with
non-relativistic bulk viscous matter, where the time-dependent bulk
viscosity has the form $\xi =\xi _{0}+\xi _{1}H+\xi _{2}\left( \frac{\dot{H}%
}{H}+H\right) $. From the cosmic scale factor we found that in case of first
limiting conditions the deceleration parameter shows the transition from
deceleration to acceleration phase in past if \ $\bar{\xi}_{0}+\bar{\xi}_{1}>%
\bar{\alpha}$, at present if $\bar{\xi}_{0}+\bar{\xi}_{1}=\bar{\alpha}$ and
in the future if $\bar{\xi}_{0}+\bar{\xi}_{1}<\bar{\alpha}$. For second
limiting conditions transition occur in the past if $\bar{\xi}_{0}+\bar{\xi}%
_{1}<\bar{\alpha}$, at present if $\bar{\xi}_{0}+\bar{\xi}_{1}=\bar{\alpha}$
and in the future if $\bar{\xi}_{0}+\bar{\xi}_{1}>\bar{\alpha}$. While in
the absence of bulk viscosity i.e. $\xi _{0}=\xi _{1}=\xi _{2}=0$, the
deceleration parameter becomes $q=\frac{1}{2}$. Hence, to describe the
late-time acceleration of the expanding universe without invoking any dark
energy component, the cosmic fluid with bulk viscosity is the most viable
candidate. The second limiting condition is not suitable for present
observational scenario, so we have considered the first limiting condition
for our analysis. Further, for constraining the model and bulk viscous
parameter we have used Hubble data and Pantheon data sets. Hence, from the
Hubble datasets, we have the best fit ranges for the model parameters are $%
\alpha =-1.03_{-0.55}^{+0.52}$, $\xi _{0}=1.54_{-0.79}^{+0.83}$, $\xi
_{1}=0.08_{-0.49}^{+0.49}$ and $\xi _{2}=0.66_{-0.83}^{+0.82}$ and from the
Pantheon datasets, we have $\alpha =-1.33_{-0.43}^{+0.45}$, $\xi
_{0}=0.10_{-0.12}^{+0.21}$, $\xi _{1}=1.81_{-0.87}^{+0.91}$ and $\xi
_{2}=2.08_{-0.96}^{+0.91}$. Furthermore, including the six data
points of the BAO datasets with the Hubble datasets and combine the results
of Hubble constrained values, we obtain the values of the model parameters
as, $\alpha =-1.06_{-0.82}^{+0.34}$, $\xi _{0}=2.25_{-1.7}^{+0.37}$%
, $\xi _{1}=0.08_{-0.96}^{+1.1}$ and $\xi
_{2}=0.7_{-1.1}^{+1.1}$. Similarly, including the six data points of
the BAO datasets with the Pantheon datasets and combine the results of
Pantheon constrained values, we obtain the values of the model parameters
as, $\alpha =-1.65_{-0.25}^{+0.85}$, $\xi _{0}=0.86_{-1.1}^{+0.29}$%
, $\xi _{1}=1.90_{-0.99}^{+0.97}$ and $\xi
_{2}=1.93_{-0.91}^{+1.2}$. For the above set of values of the model
parameters obtained are then used to plot the statefinder diagnostics.
 Finally, we conclude that the present bulk viscous model has been departed
from $\Lambda $CDM point, in the present scenario it lie in quintessence
region and it will again pass through the $\Lambda $CDM fixed point and
hence our model is different from other models of the dark energy. We
conclude that the bulk viscous theory can be considered as an alternate
theory to describe the late time acceleration of the universe.

\section*{Acknowledgments}

RS acknowledges University Grants Commission (UGC), New Delhi, India for
awarding Junior Research Fellowship (UGC-Ref. No.: 191620096030). PKS
acknowledges CSIR, New Delhi, India for financial support to carry out the
Research project [No.03(1454)/19/EMR-II Dt.02/08/2019]. We are very much grateful to the honorable referee and to the editor for the
illuminating suggestions that have significantly improved our work in terms
of research quality, and presentation.


\end{document}